\documentclass[12pt,preprint,amsmath]{aastex}

\shorttitle{Radio Flare from XRF\,050416a}
\shortauthors{Soderberg et al.}

\def\grb{XRF\,050416a}

\def\cit{1}
\def\ociw{6}
\def\prince{7}
\def\hubble{8}
\def\nrao{4}
\def\uh{9}
\def\anu{10}
\def\psu{5}
\def\srl{3}
\def\tapir{2}
\def\uva{11}
\def\ut{12}
\def\ias{13}
\def\llnl{14}

\begin{document}

\title{A Spectacular Radio Flare from XRF\,050416a at 40 days and Implications for the Nature of X-ray Flashes}

\author{
A.~M.~Soderberg\altaffilmark{\cit},
E.~Nakar\altaffilmark{\tapir},
S.~B.~Cenko\altaffilmark{\srl},
P.~B.~Cameron\altaffilmark{\cit},
D.~A.~Frail\altaffilmark{\nrao},
S.~R.~Kulkarni\altaffilmark{\cit},
D.~B.~Fox\altaffilmark{\psu},
E.~Berger\altaffilmark{\ociw,}\altaffilmark{\prince,}\altaffilmark{\hubble},
A.~Gal-Yam\altaffilmark{\cit,}\altaffilmark{\hubble},
D-S.~Moon\altaffilmark{\srl},
P.~A.~Price\altaffilmark{\uh},
G.~Anderson\altaffilmark{\anu},
B.~P.~Schmidt\altaffilmark{\anu},
M.~Salvo\altaffilmark{\anu},
J.~Rich\altaffilmark{\anu},
A.~Rau\altaffilmark{\cit},
E.~O.~Ofek\altaffilmark{\cit},
R.~A. Chevalier\altaffilmark{\uva},
M. Hamuy\altaffilmark{\ociw},
F.~A. Harrison\altaffilmark{\srl},
P. Kumar\altaffilmark{\ut},
A. MacFadyen\altaffilmark{\ias},
P.~J. McCarthy\altaffilmark{\ociw},
H.~S. Park\altaffilmark{\llnl},
B.~A. Peterson\altaffilmark{\anu},
M.~M. Phillips\altaffilmark{\ociw},
M. Rauch\altaffilmark{\ociw},
M. Roth\altaffilmark{\ociw},
S. Shectman\altaffilmark{\ociw}
}

\altaffiltext{\cit}{Division of Physics, Mathematics and Astronomy,
105-24, California Institute of Technology, Pasadena, CA 91125}

\altaffiltext{\tapir}{Theoretical Astrophysics, 130-33
        California Institute of Technology, Pasadena, CA 91125}

\altaffiltext{\srl}{Space Radiation Laboratory, MS 220-47, California 
Institute of Technology, Pasadena, CA 91125}

\altaffiltext{\nrao}{National Radio Astronomy Observatory, Socorro,
NM 87801}

\altaffiltext{\psu}{Department of Astronomy and Astrophysics, 
Pennsylvania State University, 525 Davey Laboratory, University 
Park, PA 16802}

\altaffiltext{\ociw}{Observatories of the Carnegie Institution
of Washington, 813 Santa Barbara Street, Pasadena, CA 91101}
 
\altaffiltext{\prince}{Princeton University Observatory,
Peyton Hall, Ivy Lane, Princeton, NJ 08544}
 
\altaffiltext{\hubble}{Hubble Fellow}

\altaffiltext{\uh}{Institute for Astronomy, University of Hawaii, 
2680 Woodlawn Drive, Honolulu, HI 96822}

\altaffiltext{\anu}{Research School of Astronomy and Astrophysics, 
Australian National University, Mt Stromlo Observatory, via Cotter Rd,
Weston Creek, ACT 2611, Australia}

\altaffiltext{\uva}{Department of Astronomy, University of Virginia, P.O. Box 3818, Charlottesville, VA 22903-0818}

\altaffiltext{\ut}{Astronomy Department, University of Texas, Austin, TX 78731}

\altaffiltext{\ias}{Institute for Advanced Study, Princeton, NJ 08540}

\altaffiltext{\llnl}{Lawrence Livermore National Laboratory, 7000 East Avenue, Livermore, CA 94550}

\begin{abstract}
We present detailed optical, near-infrared, and radio observations of
the X-ray flash 050416a obtained with Palomar and Siding Springs
Observatories as well as the {\it Hubble Space Telescope} and Very
Large Array, placing this event among the best-studied X-ray flashes
to date.  In addition, we present an optical spectrum from Keck LRIS
from which we measure the redshift of the burst, $z=0.6528$.  At this
redshift the isotropic-equivalent prompt energy release was about
$10^{51}$ erg, and using a standard afterglow synchrotron model we
find that the blastwave kinetic energy is a factor of 10 larger,
$E_{K,{\rm iso}}\approx 10^{52}$ erg.  The lack of an observed jet
break to $t\sim 20$ days indicates that the opening angle is
$\theta_j\gtrsim 7^\circ$ and the total beaming-corrected relativistic
energy is $\gtrsim 10^{50}$ erg.  We further show that the burst
produced a strong radio flare at $t\sim 40$ days accompanied by an
observed flattening in the X-ray band which we attribute to an abrupt
circumburst density jump or an episode of energy injection (either
from a refreshed shock or off-axis ejecta).  Late-time observations
with the {\it Hubble Space Telescope} show evidence for an associated
supernova with peak optical luminosity roughly comparable to that of
SN\,1998bw.  Next, we show that the host galaxy of XRF\,050416a is
actively forming stars at a rate of at least 2 M$_\odot$ yr$^{-1}$
with a luminosity of $L_B\approx 0.5L^{*}$ and metallicity of $Z\sim
0.2-0.8$ $Z_{\odot}$.  Finally, we discuss the nature of \grb\ in the
context of short-hard gamma-ray bursts and under the framework of
off-axis and dirty fireball models for X-ray flashes.

\end{abstract}


\section{Introduction}

Nearly a decade ago, X-ray flashes (XRFs) were observationally
recognized as a subclass within the sample of gamma-ray bursts (GRBs)
detected by the {\it BeppoSAX} Wide-Field Cameras \citep{hik+01}.  The
events are distinguished by a prompt spectrum that peaks in the
soft X-ray range ($E_p\lesssim 25$ keV), a factor of $\sim 10$ below
the typical values observed for GRBs \citep{bmf+93}.  Since then, it
has been shown that XRFs and GRBs share many observational properties,
including prompt emission durations \citep{slk+05}, redshifts
\citep{skb+04a}, broadband afterglows (e.g., XRF\,050406,
\citealt{rmb+06,smo+06}), and host galaxy properties
\citep{bfv+03,jhf+04,rsg05}.  Moreover, the recent discovery of Type
Ic supernovae (SNe) in association with XRFs 020903
\citep{skf+05,bfs+06} and 060218 \citep{pmm+06,mha+06,msg+06} indicate that
XRFs, like GRBs, are produced in massive stellar explosions.
Together, these clues strongly suggest that XRFs and GRBs share
similar progenitors.

Driven by this progress, several theories have been proposed to
explain the soft prompt emission observed for XRFs under the framework
of a standard GRB model.  One popular idea posits that XRFs are merely
typical GRBs viewed away from the collimation axis (e.g.,
\citealt{yyn+03}).  In this scenario the prompt emission is primarily
beamed away from our line-of-sight, resulting in lower fluence and
$E_p$ values for the observed burst.  An important implication of the
off-axis model is that the early afterglow evolution should be
characterized by a rising phase as the jet decelerates and spreads
sideways into our line-of-sight \citep{gpk+02,wax04,snb+06}.

Another theory suggests that XRFs are intrinsically different from
GRBs in their ability to couple energy to highly-relativistic
material.  In this scenario, XRFs are produced in explosions
characterized by lower bulk Lorentz factors, $10 \lesssim \Gamma
\lesssim 100$, than those inferred for typical GRBs, $\Gamma\gtrsim
100$ \citep{zwh04}.  This may be the result of baryon loading of the
high-velocity ejecta, a so-called ``dirty fireball'' \citep{dcb99}.
Generally speaking, low Lorentz factor explosions may be identified
through an analysis of their prompt emission since an optically thin
spectrum at high energies implies a lower limit on the Lorentz factor
\citep{ls01}.  In the case of X-ray flashes, however, there are
generally insufficient high energy photons for this type of analysis.
For these events, detailed modeling of the broadband afterglow may be
used to place a lower limit on the Lorentz factor.

Here we present an extensive, multi-frequency data set for
XRF\,050416a at $z=0.6528$ which extends to $t\sim 220$ days after the
burst.  By combining near-infrared, optical, ultra-violet, radio and
X-ray data we present an in-depth analysis of the afterglow,
energetics, supernova and host galaxy of \grb, placing it among the
best-studied X-ray flashes to date.  Moreover, thanks to our dedicated
late-time monitoring campaign, we show that \grb\ produced a strong
radio flare at $t\sim 40$ days accompanied by a brief plateau phase in
the X-ray band.  Finally, we discuss the nature of \grb\ in the
context of off-axis and dirty fireball models for X-ray flashes.

\section{Observations}
\label{Sec:obs}
XRF\,050416a was discovered by the {\it Swift} Burst Alert Telescope
(BAT) on 2005 April 16.4616 UT.  The ratio of $15-25$ keV and $25-50$
keV channel fluences, $f_{15-25~\rm keV}/f_{25-50~\rm keV}\approx
1.1$, classifies the event as an X-ray flash \citep{sbb+06}.  This is
consistent with the low peak photon energy, $E_p=15.0^{+2.3}_{-2.7}$
keV \citep{sbb+06}; a factor of $\sim 10$ lower than the typical
values observed for long-duration GRBs \citep{bmf+93} and a factor of
$\sim 3$ larger than the values inferred for XRFs 020903 and 060218
\citep{slg+04,cmb+06}

As discussed by \citet{sbb+06}, the prompt emission light-curve is
characterized by a relatively smooth, triangular peak which is only
detected at energies below 50 keV.  The burst duration is
$T_{90}\approx 2.4$ sec (15-150 keV), placing it between the classes
of short- and long-duration GRBs while the hardness ratio shows a
clear softening.  In addition, \citet{sbb+06} note two intriguing
features of the data: (1) the rise time of the pulse is longer than
the decay time, and (2) the cross-correlation lag function (an
indication of the spectral softening) is $-0.066^{+0.014}_{-0.018}$
sec, apparently inconsistent with the overall softening trend observed
for the light-curves.  This lag function estimate is
significantly different than the typical values inferred for
long-duration GRBs and even more extreme than the zero spectral lags
inferred for short-hard bursts \citep{nb06}.  We note, however, that
the spectral lag for \grb\ was derived through a comparison of the two
softest BAT bands (15-25 and 25-50 keV) and therefore prevents a clear
comparison with the spectral lag estimates for other {\it Swift} GRBs
for which 15-25 and 50-100 keV bands are typically used.  Given that
the temporal evolution is not strongly variable, this may indicate
that the prompt emission was produced by another process
(e.g. external shocks, \citealt{dcb99}).  Finally, we note that this
burst is inconsistent with the lag-luminosity correlation for
long-duration GRBs which posits that low luminosity bursts such as
\grb\ have long spectral lags \citep{nmb00}.

\subsection{Early Optical Observations}
\label{sec:optical}
\begin{figure}
\vspace{-1.75cm}
\plotone{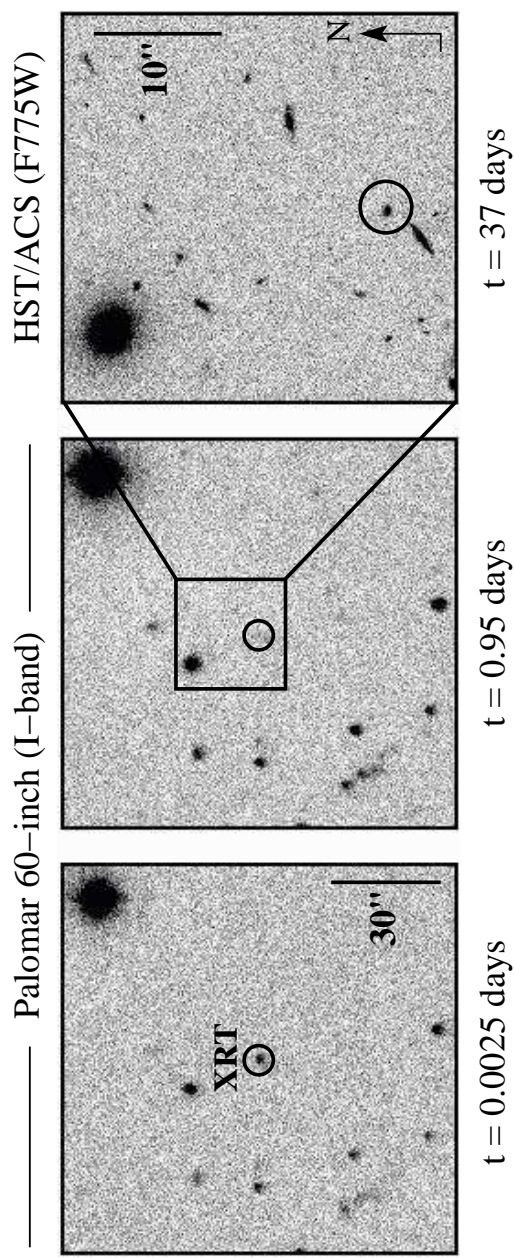}
\vspace{-1.75cm}
\caption{Discovery image of the afterglow of XRF\,050416a.  We began
  observing the field of XRF\,050416a with the Palomar 60-inch
  telescope in the $I-$band about 2.5 minutes after the burst.  We
  discovered a fading source within the 3.0 arcmin (radius) {\it
    Swift}/BAT error circle.  Subsequent localizations of the XRT
  (circle) and UVOT afterglow positions were shown to be coincident
  with the P60 source. Our late-time {\it HST} images reveal a
  host galaxy coincident with the optical afterglow position.}
\label{fig:field}
\end{figure}

Using the roboticized Palomar 60-inch telescope (P60) we
initiated observations of the field of XRF\,050416a at 2005 April
16.4634 UT (2.5 minutes after the burst).  In our first 120
second image we discovered a new source within the {\it Swift}/BAT
error circle at $\alpha=12^{\rm h} 33^{\rm m} 54^{\rm
s}.58$, $\delta=+21^{\circ} 03' 26''.7$ (J2000) with an uncertainty of
0.5 arcsec in each coordinate based on an astrometric tie to the
USNO-B catalog (Figure~\ref{fig:field}).  We subsequently monitored
the afterglow evolution with the Palomar 60-inch, 200-inch, and Siding
Springs 2.3-meter telescopes in the $R-$, $I-$, $z'-$ and $K_s-$band
through $t\approx 7$ days.

\begin{figure}
\epsscale{1.0}
\plotone{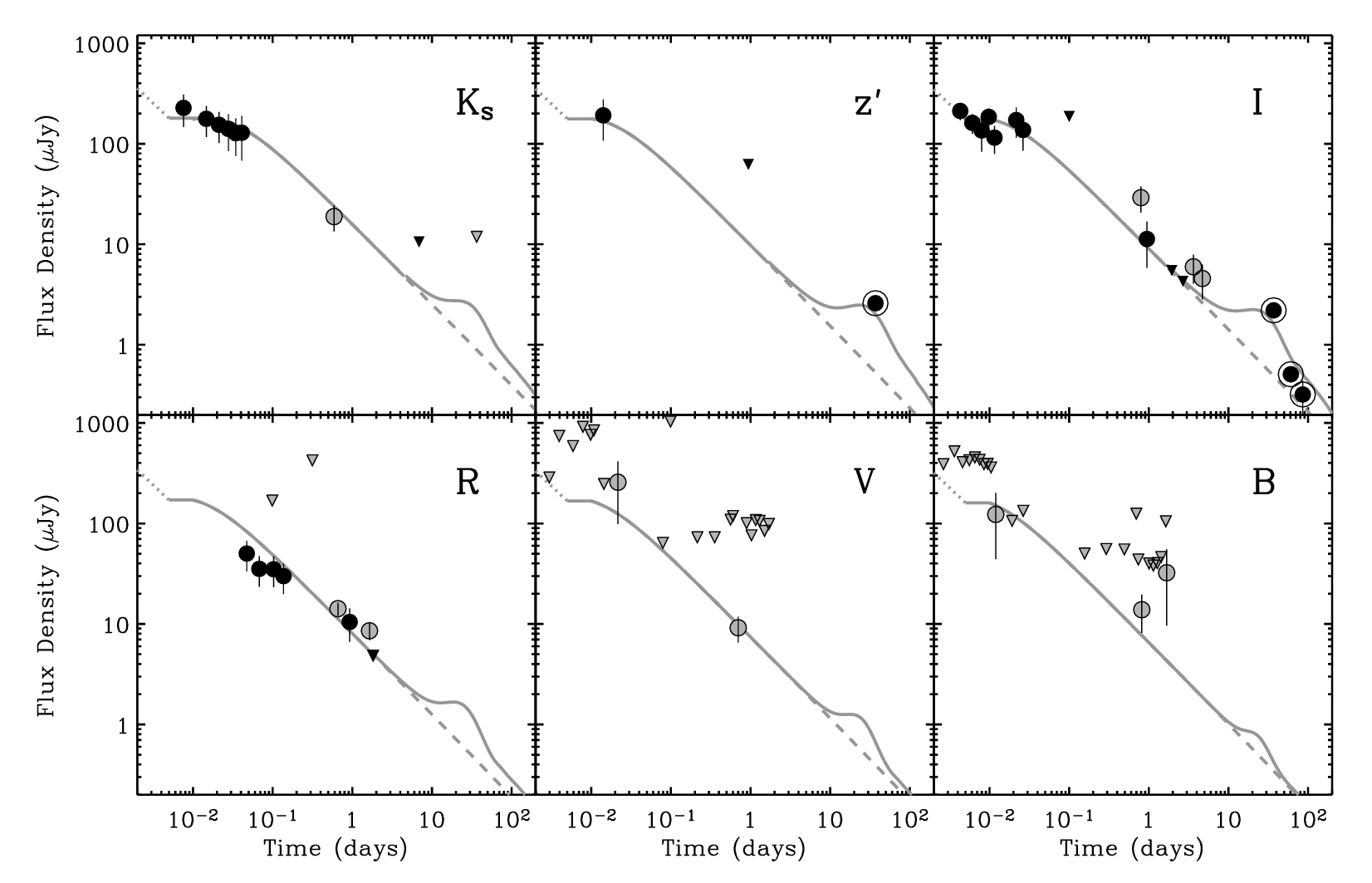}
\caption{Optical ($BVRIz'$) and NIR ($K_s$) light-curves of the
  afterglow of XRF\,050416a.  We supplement our measurements from
  Table~\ref{tab:NIRopt} (black symbols) with data from the GCNs
  \citep{lcj+05,pmc+05,qll+05,tor05,ytk05} and \citet{hbg+06} (grey
  symbols).  Detections are shown as circles and upper limits as
  inverted triangles.  $R-$band data from the ANU 2.3-meter have been
  binned for clarity.  All data have been corrected for host galaxy
  extinction (see \S\ref{sec:optical}). Between $t\approx 0.01$ and 1
  day the NIR/optical afterglow shows an average decay index of
  $\overline{\alpha}_{\rm NIR/opt}\approx -0.75$.  As described in
  \S\ref{sec:fit}, we find a reasonable fit to the spectral and
  temporal evolution of the broadband data with a standard afterglow
  model.  Our {\it HST} measurements are shown as
  encircled dots; the implied steep spectral index suggests the
  contribution from an supernova component at $t\sim 40$ days
  (\S\ref{sec:sn}). We sum the flux from a SN\,1998bw-like supernova
  at $z=0.6528$ with our afterglow model to produce the final fits
  (grey solid lines). For comparison, we also show the afterglow model
  alone (dashed grey line).  We note that this model does not apply to
  the very early afterglow evolution at $t\lesssim 0.01$ days.  The
  dotted grey lines at early time represent the evolution of the
  NIR/optical assuming these bands track the X-ray evolution on this
  timescale.}
\label{fig:NIRopt_model_ltcurves2}
\end{figure}

Aperture photometry was performed on each of the images in the
standard method using the {\tt apphot} package within IRAF.  Absolute
calibration of $R$-, $I$-, and $z'-$band light-curves was derived
using the field calibration of \citet{hen05} and utilizing the
transformation equations of \citet{stk+02}.  The $K_s$-band
light-curve was calibrated against 2MASS using 15 unconfused sources.
The errors resulting from calibration uncertainty ($\lesssim 10\%$) were
added in quadrature to the measurement errors.  As shown in 
Figure~\ref{fig:NIRopt_model_ltcurves2} and Table~\ref{tab:NIRopt},
the afterglow was $I=18.82\pm 0.11$ mag at $t\approx 1.6$ min
(mid-exposure).

We supplement these NIR/optical afterglow data with additional
measurements from the GCNs \citep{lcj+05,pmc+05,qll+05,tor05,ytk05}
and those reported by \citet{hbg+06}, obtained with the {\it
Swift}/UVOT and 1.54-m Danish Telescope.  The resulting dataset spans
1930\AA\ ({\it Swift}/UVOT UVW2) to 22000\AA\ ($K_s$); however, we
note that the majority of the UV observations are upper limits.  Using
this extended dataset we measure the temporal and spectral properties
of the NIR/optical afterglow emission.  We find the following
NIR/optical power-law decay indices ($\alpha$, where $F_{\nu}\propto
t^{\alpha}$) between $\sim 0.01$ and 1 day: $\alpha_{K_s}=-0.7\pm
0.3$, $\alpha_I=-0.7\pm 0.3$, $\alpha_R=-0.5\pm 0.3$,
$\alpha_V=-1.0\pm 0.3$, $\alpha_B=-0.5\pm 0.3$, consistent with the
values reported by \citet{hbg+06}.  These values imply a mean temporal
index of $\overline{\alpha}_{\rm NIR/opt}=-0.7\pm 0.2$.

Finally, we analyze the spectral index ($\beta$ with $F_{\nu}\propto
\nu^{\beta}$) within the NIR/optical bands.  As shown in
Figure~\ref{fig:spectrum} there are two epochs at which the
photometric spectrum is well-sampled: $t\approx 0.014$ and 0.8
days. We fit each of the observed spectra with a simple power-law and
find $\beta_{\rm NIR/opt}\approx -1.3$ ($\chi^2_r\approx 0.5$) and $-1.5$
($\chi^2_r\approx 0.6$) for the first and second epochs, respectively.
As will be discussed in \S\ref{sec:prelim}, the observed steep
spectrum is indicative of extinction within the host galaxy.

\subsection{Late-time Observations with HST}
\label{sec:hst}
Using the {\it Wide-Field Camera} ({\it WFC}) of the {\it Advanced
  Camera for Surveys} ({\it ACS}) on-board the {\it Hubble Space
  Telescope} ({\it HST}), we imaged the field of XRF\,050416a four
times, spanning 37 to 219 days after the burst.  Each epoch
consisted of two or four orbits during which we imaged the field in
filters F775W and/or F850LP, corresponding to SDSS $i'$- and $z'$-
bands, respectively.

The {\it HST} data were processed using the {\tt multidrizzle} routine
\citep{fh02} within the {\tt stsdas} package of IRAF.  Images were
drizzled using {\tt pixfrac}=0.8 and {\tt pixscale}=1.0 resulting in a
final pixel scale of 0.05 arcsec/pixel.  Drizzled images were then
registered to the final epoch using the {\tt xregister} package within
IRAF.  We astrometrically tied the {\it HST} and P60 images
using 12 unconfused sources in common resulting in a final systematic
uncertainty of 0.70 arcsec ($2\sigma$).

To search for source variability and remove host galaxy contamination,
we used the {\it ISIS} subtraction routine by \citet{a00} which
accounts for temporal variations in the stellar PSF.  Adopting the
final epoch observations as templates we produced residual
images. These residual images were examined for positive sources
positionally coincident with the P60 afterglow.

Photometry was performed on the residual sources within a 0.5 arcsec
aperture.  We converted the photometric measurements to infinite
aperture and calculated the corresponding AB magnitudes within the
native {\it HST} filters using the aperture corrections and
zero-points provided by \citet{sjb+05}.  Here we made the reasonable
assumption that the transient flux is negligible in the template images.
For comparison with ground-based data, we also converted the F775W
measurements to Johnson $I$-band (Vega) magnitudes using the
transformation coefficients derived by \citet{sjb+05} and assuming the
$F_{\nu}$ source spectrum implied by the first epoch {\it HST} data.

\begin{figure}
\vspace{-1cm}
\epsscale{1.0}
\plotone{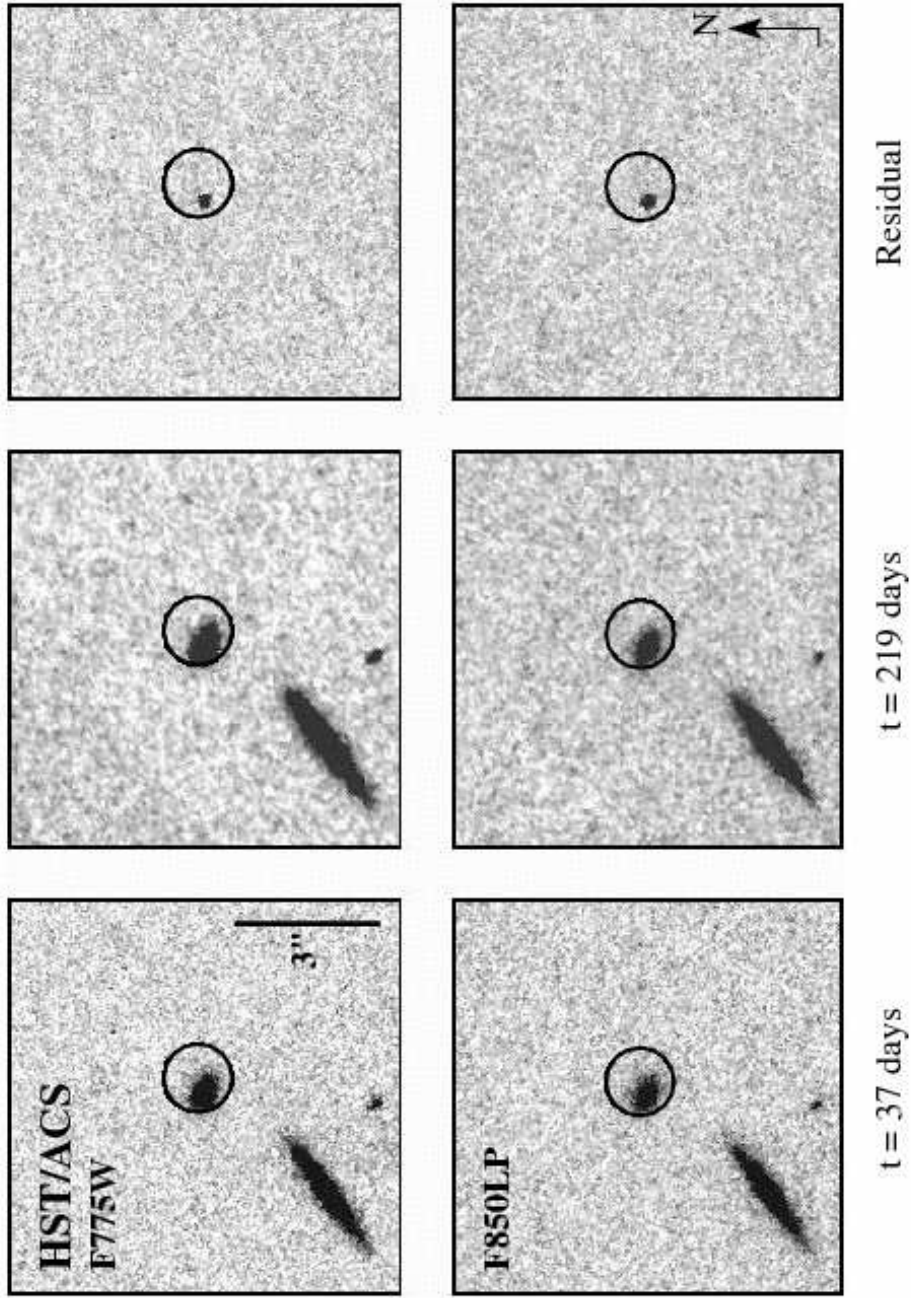}
\vspace{-1.5cm}
\caption{{\it HST} images of XRF\,050416a were obtained with {\it ACS}
  in filters F775W and F850LP spanning $t\approx 37$ to 219 days after
  the explosion.  The host galaxy emission dominates that of the
  afterglow throughout this timescale.  Relative astrometry between
  the P60 and the {\it HST} images provides an optical afterglow
  position accurate to 0.70 arcsec (circle, $2\sigma$).  Residual
  images of the afterglow were produced through image subtraction
  techniques (\S\ref{sec:hst}).  The position of the afterglow is
  offset by $0.02\pm 0.02$ arcsec with respect to the host galaxy
  center.}
\label{fig:images_full}
\end{figure}

As shown in Table~\ref{tab:hst} and Figure~\ref{fig:images_full}, the
transient is clearly detected in the first epoch {\it HST} residual
images.  An astrometric tie between the first and final epochs shows
that the residual is offset $0.02\pm 0.02$ arcsec from the center of
the host galaxy.  Our residual images show the source to be
F775W$=24.35\pm 0.02$ and F850LP$= 23.83\pm 0.03$ mag in
the AB system ($I=23.82\pm 0.02$ mag in the Vega system) at
$t\approx 37$ days.  As shown in
Figure~\ref{fig:NIRopt_model_ltcurves2}, these values are a factor of
$\sim 5$ above an extrapolation of the early afterglow decay.  The
observed spectral index between the F775W and F850LP filters 
is $\beta_{HST}\approx -2.8\pm 0.3$, significantly steeper than the
afterglow spectrum observed at early time (\S\ref{sec:optical} and
Figure~\ref{fig:spectrum}) as well as the typical values measured for
NIR/optical afterglows ($\beta_{\rm NIR/opt}\approx -0.6$;
\citealt{pk02,yhs+03}).  As will be discussed in \S\ref{sec:sn}, the
timescale and spectral signature of the observed flux excess are
suggestive of a thermal supernova component.  

\subsection{Spectroscopic Observations}
\label{sec:label}

We observed the host galaxy of XRF\,050416a with the Low Resolution 
Imaging Spectrograph on Keck I on 2005 June 6.3 UT ($t\sim 50$ days).
We placed a 1.0 arcsec longslit across the host galaxy at a position angle of
PA=$87^{\circ}$.  Data were reduced in standard manner using the {\tt
  onedspec} and {\tt twodspec} packages within IRAF.  Flux calibration
was performed using the spectrophotometric standard star BD+284211.

\begin{figure}
\plotone{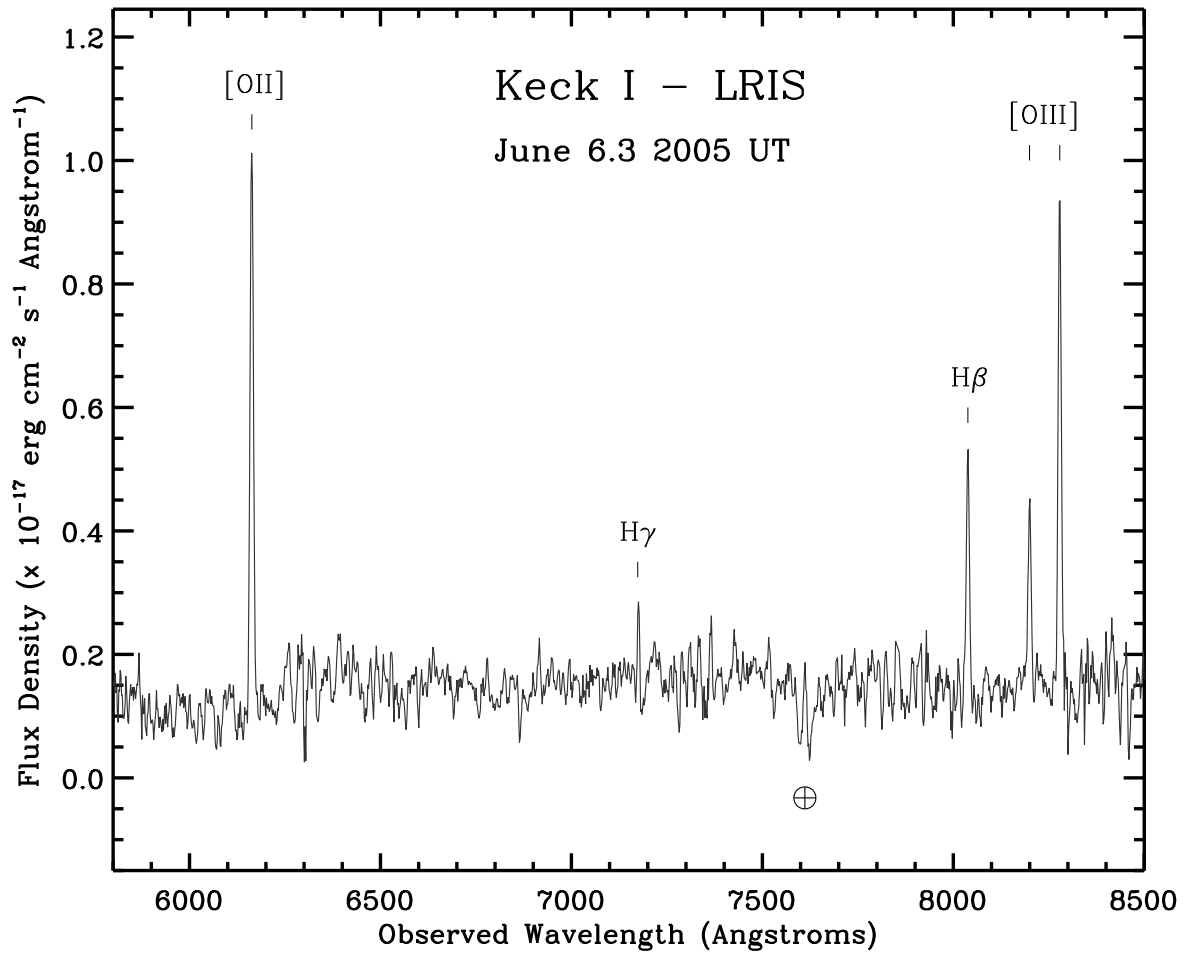}
\caption{Optical spectrum of the host galaxy of XRF\,050416a as
  observed with the Low Resolution Imaging Spectrograph on Keck I on
  2005 June 6.3 UT.  Several bright emission lines are detected,
  indicating that the host is a star-forming galaxy at redshift
  $z=0.6528\pm 0.0002$.  As discussed in \S\ref{sec:host} we find that
  the host is actively forming stars with a rate of $\gtrsim
  2~$M$_\odot$ yr$^{-1}$ and an inferred metallicity of $Z\sim
  0.2-0.8~Z_{\odot}$.}
\label{fig:lris_spec}
\end{figure}

As shown in Figure~\ref{fig:lris_spec} and Table~\ref{tab:lines}, we
detect several strong emission lines in the spectrum including
H$\beta$, H$\gamma$, [\ion{O}{2}]$\lambda 3727$, and
[\ion{O}{3}]$\lambda\lambda 4959,5006$ at a redshift of $z=0.6528\pm
0.0002$.  Adopting the standard cosmological parameters ($H_0=71~\rm
km~s^{-1}~Mpc^{-1}$, $\Omega_M=0.27$, $\Omega_{\Lambda}=0.73$), the
isotropic gamma-ray energy release is $E_{\gamma,\rm iso}\approx
1.2\pm 0.2\times 10^{51}$ erg (1 keV -- 10 MeV; \citealt{sbb+06}).
Compared with typical long-duration bursts, the prompt energy release
of \grb\ is a factor of $\sim 100$ lower (\citealt{fks+01,bfk03,ama06}
and references therein).

\subsection{Radio Observations}
\label{sec:radio}

We began observing XRF\,050416a with the Very Large Array\footnote{The
  Very Large Array and Very Long Baseline Array are operated by the
  National Radio Astronomy Observatory, a facility of the National
  Science Foundation operated under cooperative agreement by
  Associated Universities, Inc.} (VLA) on 2005 April 16.49 UT
  ($t\approx 37$ min).  No radio source was detected coincident with
  the optical position to a limit of $F_{\nu} < 122~ \mu$Jy at
  $\nu=8.46$ GHz.  However, further observations at $t\approx 5.6$
  days revealed a new radio source with $F_{\nu}\approx 101\pm
  34~\mu$Jy coincident with the optical and X-ray afterglow positions
  at $\alpha=12^{\rm h} 33^{\rm m} 54.594^{\rm s}\pm 0.002$,
  $\delta=21^{\circ} 03' 26.27''\pm 0.04$ (J2000) which we identify as
  the radio afterglow.

We continued to monitor the radio afterglow at 1.43, 4.86 and 8.46 GHz
through $t\approx 140$ days (Table~\ref{tab:vla}).  All observations were
taken in standard continuum observing mode with a bandwidth of
$2\times 50$ MHz. We used 3C286 (J1331+305) for flux calibration,
while phase referencing was performed against calibrators J1221+282 and
J1224+213.  The data were reduced using standard packages within the
Astronomical Image Processing System (AIPS).  

As shown in Figure~\ref{fig:vla_lt_curve}, the evolution of the radio
afterglow is dissimilar from those of typical GRBs and inconsistent
with a standard blastwave model.  Between $t\sim 20$ and 40 days an
abrupt rebrightening (factor of $\sim 3$) is observed at all radio
bands with a temporal index steeper than $\alpha_{\rm rad}\approx
0.9$.  Following this radio flare the emission decays rapidly with an
index of $\alpha_{\rm rad}\lesssim -1.5$.  The peak radio luminosity
at 8.46 GHz, $L_{\rm rad}\approx 1.3\times 10^{31}~\rm
erg~s^{-1}~Hz^{-1}$, is typical for GRBs \citep{fkb+03}, a factor of
$\sim 10$ and $10^3$ larger than those of XRFs 020903 and 060218,
respectively \citep{skb+04a,skn+06}, and between $10^2$ and $10^6$
times higher than the peak radio luminosities observed for
optically-selected Type Ibc supernovae \citep{sod06}.  By $t\approx
105$ days the radio afterglow is no longer detected at any frequency.

\subsection{X-ray Observations}
\label{sec:xray}

\begin{figure}
\plotone{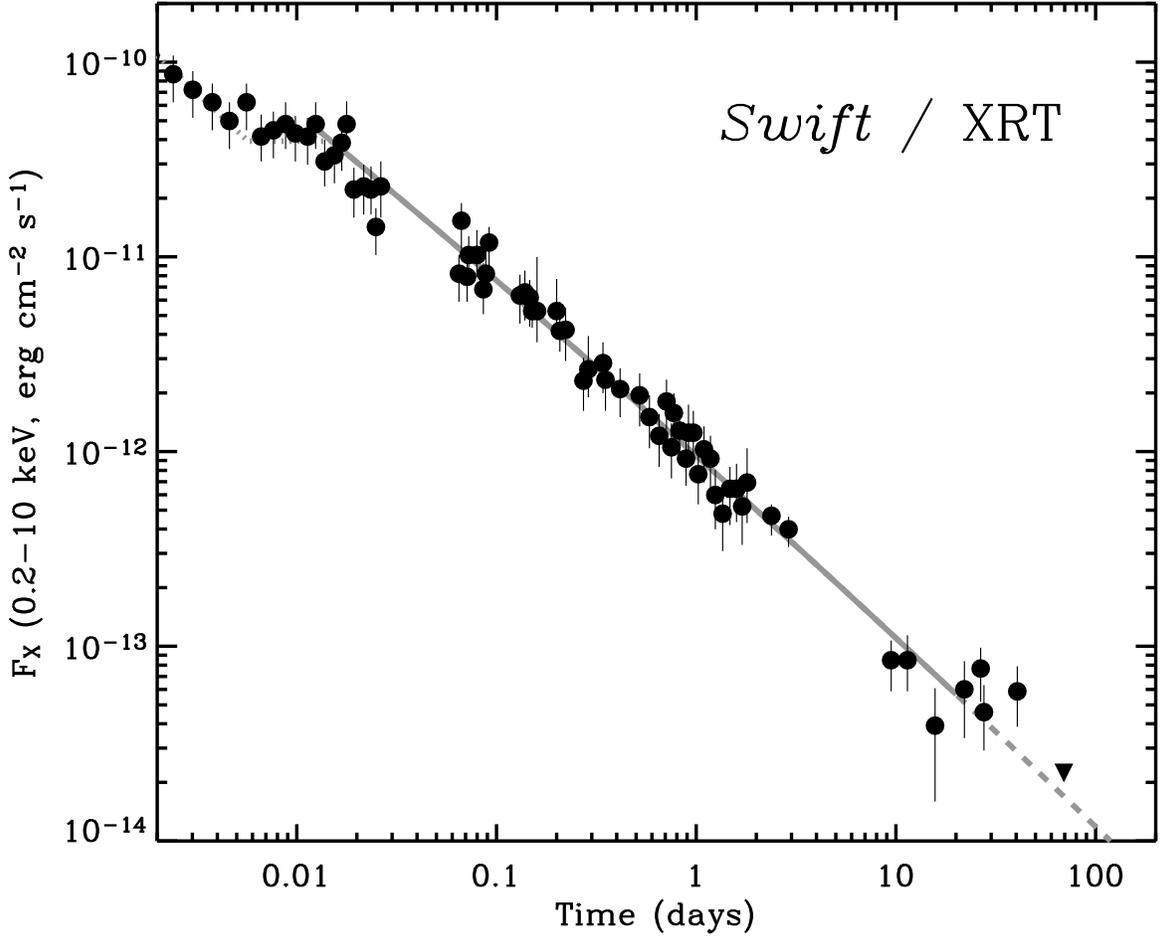}
\caption{X-ray observations of \grb\ from the {\it Swift}/XRT as
  reported by \citet{mlc+06}.  The X-ray light-curve is characterized
  by four phases: (1) a initial steep decay, (2) a flattening between
  $t\approx 0.005-0.01$ days, (3) a subsequent decay, and (4) a second
  flattening between $t\approx 20-40$ days.  The final measurement at
  70 days implies a rapid steepening following the second flattening.
  We attribute the late-time flattening to energy injection
  (refreshing shell or off-axis ejecta) or a large circumburst density
  jump.  Overplotted is our afterglow model fit between 0.014 and 20
  days (solid grey line) and extrapolated to late time (dashed grey
  line).  We do not attempt to fit the data prior to 0.01 days (dotted
  line) due to insufficient broadband data.}
\label{fig:xrt_lt_curve}
\end{figure}

The afterglow of XRF\,050416a was observed with the {\it Swift} X-ray
Telescope (XRT) beginning $1.2$ minutes after the burst and continuing
through $t\sim 69$ days, placing it among the best studied X-ray
afterglows to date \citep{nkg+06,owo+06}.  From their analysis of the
XRT data, \citet{mlc+06} report a spectral index of
$\beta_X=-1.04^{+0.11}_{-0.05}$ and evidence for significant
absorption in the host galaxy, $N_H=6.8^{+1.0}_{-1.2}\times
10^{21}~\rm cm^{-2}$, corresponding to $E(B-V)_{\rm rest}=1.2\pm 0.2$
assuming the conversion of \citet{ps95} and a Milky Way extinction
curve \citep{pei92}.  As shown in Figure~\ref{fig:xrt_lt_curve}, at
$t=10$ hours the X-ray luminosity was $L_{\rm X,iso}(t=10~{\rm
hrs})\approx 2.3\times 10^{45}~\rm erg~s^{-1}$, placing it at the
lower edge of the observed distribution for GRBs \citep{bkf03,fw01}.

As discussed by \citet{mlc+06} and shown in
Figure~\ref{fig:xrt_lt_curve}, the early evolution of the X-ray
afterglow can be characterized by three phases: (1) an initial steep
decay (2) a flattening between $t\approx 7$ and 20 minutes during
which the flux is roughly constant, and (3) a resumed decay through
$t\sim 20$ days.  These three phases have been shown to be ubiquitous
among {\it Swift} X-ray afterglows \citep{nkg+06}.  However, at $t\sim
20$ days there is a second flattening which continues through $t\sim
40$ days.  By $70$ days the X-ray afterglow is no longer detected,
implying a significant steepening to $\alpha_X\lesssim -1.8$ between
the last two observations.  We note that the timescale for the X-ray
flattening and subsequent steep decay is coincident with the observed
radio flare.

\section{Properties of the Early Afterglow}
\label{sec:afterglow}

Using the detailed multi-frequency observations of the \grb\ afterglow
we can constrain the physical properties of the ejecta and the circumburst
density.  We adopt a standard relativistic blastwave model in which
the afterglow emission is produced through the dynamical interaction
of the ejecta with the surrounding medium (the forward shock, FS) with
an additional component from shock heating of the ejecta (the reverse
shock, RS).  In this scenario, the total energy density is partitioned
between relativistic electrons, $\epsilon_e$, and magnetic fields,
$\epsilon_B$, while the thermal energy of the shocked protons accounts
for the fraction remaining (see \citealt{pir99} for a review). The
shocked electrons are accelerated into a power-law distribution,
$N(\gamma)\propto \gamma^{-p}$ above a minimum Lorentz factor,
$\gamma_m$.  The emission resulting from the forward and reverse shock
components is described by a synchrotron spectrum characterized by
three break frequencies --- the self-absorption frequency, $\nu_a$,
the characteristic frequency, $\nu_m$, and the cooling frequency,
$\nu_c$ --- and a flux normalization, $F_{\nu_m}$ \citep{spn98}.  In
modeling the afterglow spectral and temporal evolution, we adopt the
formalism of \citet{gs02} for a relativistic forward shock expanding
into a constant density circumburst medium.

\subsection{Preliminary Constraints}
\label{sec:prelim}

In fitting the forward shock model to the afterglow data of \grb\ we
use only observations between 0.014 and 20 days when the afterglow
follows a simple power-law evolution.  
To constrain the spectrum of the forward shock, we first investigate
the afterglow evolution in the optical and X-ray bands.  As shown in
Figure~\ref{fig:xrt_lt_curve}, the X-ray data between 0.014 and 20
days are reasonably fit with $\alpha_X\approx -1.1$ ($\chi^2_r\approx
0.70$).  \citet{mlc+06} report that the X-ray spectral index on this same
timescale is $\beta_X\approx -1.04$, leading to
$\alpha-3\beta/2\approx 0.5$.  A comparison to the standard closure
relations, $\alpha-3\beta/2=0$ ($\nu_m < \nu < \nu_c$) and
$\alpha-3\beta/2=1/2$ ($\nu > \nu_c$) indicates that $\nu_X > \nu_c$.
This conclusion is supported by the near-IR to X-ray spectral slope,
$\beta_{K,X}=-0.47\pm 0.06$ at $t\approx 0.6$ days, which is flatter
than $\beta_X$ as expected if $\nu_{\rm NIR/opt} < \nu_c < \nu_X$.
Therefore, the X-ray observations suggest that $p=-2\beta\approx
2.1$.

\begin{figure}
\epsscale{0.8}
\plotone{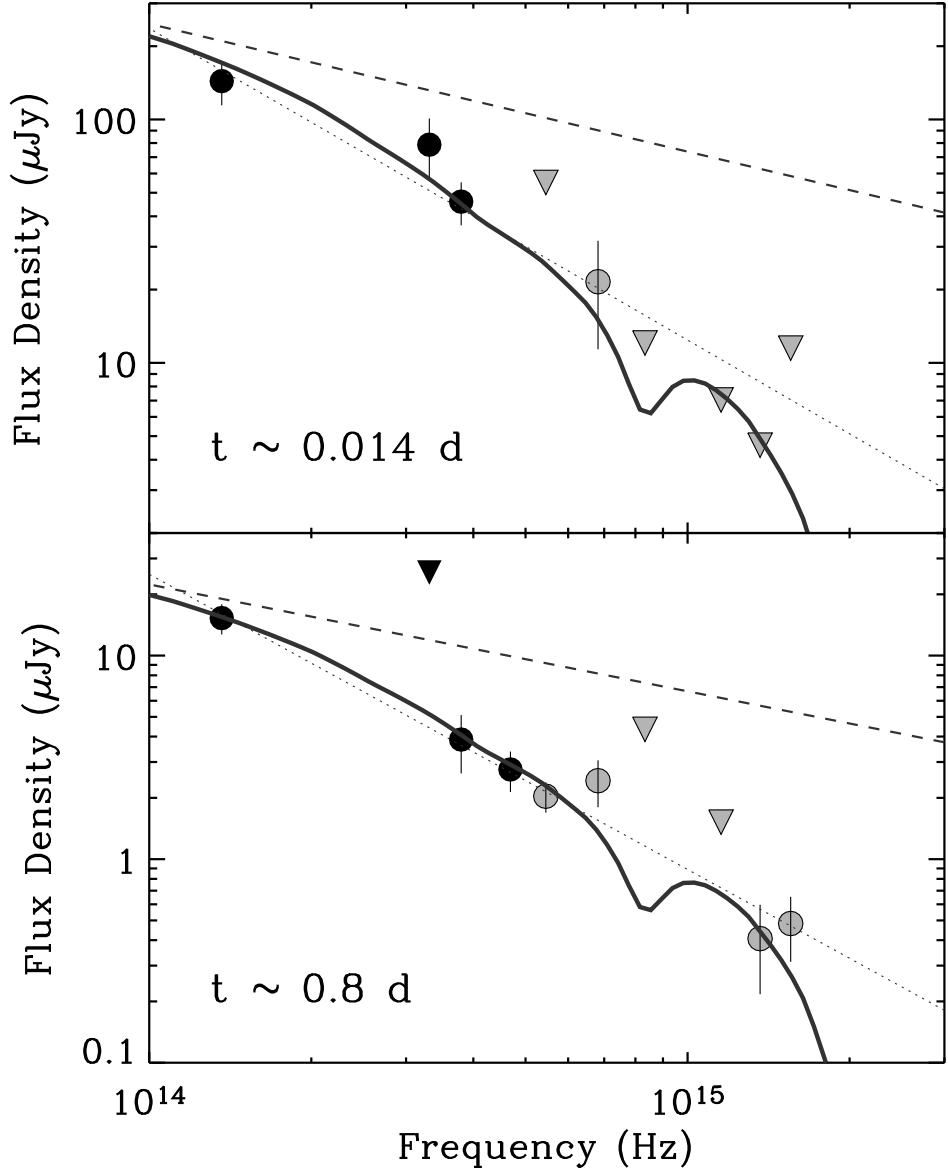}
\vspace{-1cm}
\caption{The observed photometric spectrum for XRF\,050416a is shown
  at $t\approx 0.014$ (top) and 0.8 days (bottom).  The $K_s-$, $z'-$,
  $I-$ and $R-$band data (black symbols) are from
  Table~\ref{tab:NIRopt} while the {\it V, B, U, UVW1, UVM2} and {\it
  UVW2} data (grey symbols) are from \citet{hbg+06}.  Detections are
  shown as circles and upper limits as inverted triangles. The
  observed spectrum is fit by $\beta_{\rm NIR/opt}\approx -1.3$ and
  $-1.5$ in the top and bottom panels, respectively (dotted line).  We
  fit the data with an extinction model (solid line) which includes
  $E(B-V)=0.03$ from the Galaxy \citep{sfd98} and a rest-frame host
  galaxy extinction of $E(B-V)_{\rm rest}=0.28$ assuming an intrinsic
  spectral index of $\beta_{\rm NIR/opt}\approx -0.55$ (dashed line).}
\label{fig:spectrum}
\end{figure}

Next we consider the spectral index within the NIR/optical bands.  As
discussed in \S\ref{sec:optical} and shown in
Figure~\ref{fig:spectrum}, the observed NIR/optical spectral index on
this timescale is $\beta_{\rm NIR/opt}\approx -1.3$ to $-1.5$.  These
values are significantly steeper than $\beta_{K,X}$ and imply that the
optical flux is suppressed by host galaxy extinction.  Making the
reasonable assumption that $\nu_m \lesssim \nu_{\rm NIR/opt}$ on the
timescale of our afterglow observations, and adopting $p\approx 2.1$
as indicated by the X-ray observations, we estimate $\beta_{\rm
NIR/opt}=-(p-1)/2\approx -0.55$ for the intrinsic spectral index of the
NIR/optical afterglow.  Adopting this value for $\beta_{\rm NIR/opt}$,
we find that both NIR/optical spectra are reasonably fit with a
Galactic extinction of $E(B-V)=0.03$ \citep{sfd98} and a host galaxy
component of $E(B-V)_{\rm rest}\approx 0.28$
(Figure~\ref{fig:spectrum}).  Here we have assumed a Milky Way
extinction model for the host \citep{pei92} but note that a comparable
fit may be obtained for an SMC extinction curve.  We further note that
this optically-derived extinction estimate is lower than that
inferred from the X-ray spectrum \citep{mlc+06}, consistent with the
trend observed for long-duration GRBs \citet{gw01}.

With this extinction correction, the near-IR to X-ray spectral index
becomes $\beta_{K,X}\approx -0.5\pm 0.1$, consistent with our estimate
for the intrinsic spectral index within the NIR/optical band.
Moreover, the extinction-corrected NIR/optical spectral index and
observed average temporal index of $\overline{\alpha}_{\rm
NIR/opt}=-0.7\pm 0.2$ are consistent with the standard closure
relation: $\alpha-3\beta/2=0\approx 0.1$.  We also note that this
supports our assumption of a constant density medium since in a wind
environment the expected temporal index is steeper than
$\alpha=-1.25$, and thus inconsistent with the observed values.  Using
all the available optical and X-ray observations we estimate that
$\nu_c\approx 1\times 10^{17}$ Hz at $t=1$ day.

Next we compare the near-IR and radio afterglow data to constrain
$\nu_m$ and the peak spectral flux, $F_{\nu_m}$.  Assuming that
$\nu_m$ passed through the NIR/optical bands near the time of our
first $K_s$-band observations implies that the peak spectral flux is
roughly comparable to the extinction-corrected $K_s$ flux:
$F_{K_s}\approx 230~\mu$Jy at $t\approx 11$ minutes.  Here we focus on
the $K_s$-band data since they are the least sensitive to host galaxy
extinction, which we estimate to be $A_K\approx 0.24$ mag (a $20\%$
increase in flux) for $E(B-V)_{\rm rest}\approx 0.28$.  Scaling these
constraints to $t=1$ day ($\nu_m\propto t^{-1.5}$ and
$F_{\nu_m}\propto t^0$) and accounting for the smooth shape of the
spectral peak, we find $\nu_m \approx 4.0\times 10^{11}$ Hz and
$F_{\nu_m}\approx 350~\mu Jy$.  Here and throughout, $F_{\nu_m}$ is
the asymptotic extrapolation of the smooth spectrum peak and is
therefore slightly higher than the intrinsic peak flux.  We note that
since the NIR/optical data require $F_{\nu_m}\propto \nu^{-(p-1)/2}$,
lower values of $\nu_m$ imply increasingly higher values of
$F_{\nu_m}$ at the time of the first $K_s$-band observations.

Finally, we test that these constraints are consistent with the radio
observations.  Given the evolution of $\nu_m$, these constraints
predict that the spectral peak should pass through the radio band at
$t\approx 13$ days with an extrapolated peak flux density of
$F_{\nu_m}\approx 350~\mu$Jy, roughly consistent with the 4.86 GHz
observations on this timescale.  We emphasize that the early steady
decay of the NIR/optical data require that $\nu_m$ passes through the
radio no later than 13 days.  Finally, we note that the radio
spectrum is optically thin throughout the timescale of VLA monitoring
(see Figure~\ref{fig:vla_lt_curve}) and thus we observationally
constrain $\nu_a$ to be below 1.43 GHz.

\subsection{Forward Shock Broadband Model}
\label{sec:fit}

Adopting these constraints we apply a broadband afterglow model fit to
the multi-frequency data in order to determine the physical parameters
of the burst.  The four spectral parameters ($F_{\nu_m}$, $\nu_a$,
$\nu_m$ and $\nu_c$) are fully determined by four physical parameters:
the isotropic ejecta energy, $E_{K,{\rm iso}}$, the energy density
partition fractions, $\epsilon_e$ and $\epsilon_B$, and the
circumburst density, $n$.  Therefore by constraining the four spectral
parameters through broad-band observations, we are able to determine a
unique solution for the four physical parameters (see \citealt{spn98}
and \citealt{pir99} for reviews).  Although the radio observations
provide only an upper limit on $\nu_a$, we are able to define a range
of reasonable values by requiring that $\epsilon_e$,$\epsilon_B \le
1/3$ which accounts for an equal or greater contribution from shocked
protons.  This requirement excludes unphysical solutions in which the
sum of the contributions from shocked electrons, protons and magnetic
fields exceed the total energy density.  Combined with the observed
constraints for $F_{\nu_m}$, $\nu_m$ and $\nu_c$ we find the following
ranges for the physical parameters:
\begin{mathletters}
\begin{eqnarray}
E_{K,{\rm iso}}\approx (8.2-14) \times 10^{51}~~\rm erg  \\
n\approx (0.33-4.2)\times 10^{-3}~~\rm cm^{-3}  \\
\epsilon_e \approx (0.20-1/3) \\
\epsilon_B \approx (0.072-1/3).  
\label{eqn:constr}
\end{eqnarray}
\end{mathletters}
As shown in Figures~\ref{fig:NIRopt_model_ltcurves2},
\ref{fig:vla_lt_curve}, and \ref{fig:xrt_lt_curve}, this model provides
an adequate fit to the broadband data between $t\sim 0.01$ and 20
days.

\subsection{Collimation of the Ejecta and Viewing Angle}
\label{sec:jet}

The lack of an observed jet break in the X-rays to $t\sim 20$ days,
together with the inferred physical parameters constrain the opening
angle of the jet (e.g. \citealt{sph99}) to $\theta_j\approx 3.1
t_j^{3/8} E_{{K,{\rm iso},52}}^{-1/8} n_{-3}^{1/8} (1+z)^{-3/8}\gtrsim
6.9$ degrees.  Here, $t_j$ is the jet break time in days and we have
adopted the notation $10^x Q_x=Q$. This limit is slightly larger than
the median of the jet opening angles inferred for long-duration GRBs,
$\theta_j\sim 5^{\circ}$ (\citealt{bfk03,ggl04,sbk+06} and references
therein).  This indicates that the beaming-corrected ejecta energy
release is $E_K\equiv E_{K,{\rm iso}}(1-{\rm cos}\theta)\approx
(9.8\times 10^{49} - 1.4\times 10^{52})$ erg where the range includes
the uncertainty in $E_{K,{\rm iso}}$ and the lower limit on $\theta_j$.
Moreover, we expect the blastwave to become non-relativistic on a
timescale $t_{\rm NR}\approx 2.0~E_{51}^{1/3} n_{-3}^{-1/3}\sim (0.6 -
6.8)$ yrs \citep{lw00}.  On a similar timescale, the ejecta are
predicted to approach spherical symmetry after which the blastwave
evolution is well described by the Sedov-von Neumann-Taylor (SNT)
solution \citep{snt,fwk00}; in this regime the afterglow emission
decays with $\alpha =-9/10$ ($-1$) for frequencies below (above) the
cooling frequency.

Next, the early steady decay of the X-ray and NIR/optical afterglow
indicates that the jet collimation axis is directed roughly along our
line-of-sight.  In comparison, GRBs viewed significantly off-axis
($\theta_{\rm obs} > 2\theta_j$) are predicted to show a rising
afterglow light-curve as the jet spreads sideways and intersects our
viewing angle \citep{gpk+02,wax04,snb+06}.  Here, $\theta_{\rm obs}$
is the angle between our line-of-sight and the jet collimation axis.
We conclude that the ejecta are viewed roughly on-axis and therefore
the inferred beaming-corrected energies are not affected significantly
(if at all) by viewing angle effects.

\section{Properties of the Late-time Afterglow}
\label{sec:late}

Next we address the nature of the late-time broadband afterglow
evolution with special attention to the strong radio flare observed at
$\sim 40$ days.  Radio flares have been noted for several other GRBs,
though only at early times (e.g. GRB\,990123 at $t\lesssim 1$ day;
\citealt{kfs+99}).  Based on their observed timescale and evolution,
radio flares are typically attributed to emission from the reverse
shock \citep{sp99}.  Here we present detailed radio observations for
XRF\,050416a which show for the first time a strong radio flare at
late time. Possible causes for a late-time radio rebrightening
include the emission from a decelerating jet initially directed away
from our line-of-sight \citep{wax04,ls04}, circumburst density
variations \citep{wij01,rdm+01}, and energy injection from a slow
shell catching up to and refreshing the afterglow shock \citep{rm98}.
We discuss each of these possibilities below.

\subsection{Off-axis Jet Emission}
\label{sec:offaxis}

It has been shown that the observational signature of a relativistic
jet viewed significantly away from our line-of-sight is a rapid
achromatic rise in the early afterglow light-curves
\citep{pac01,gpk+02,wax04}.  In this scenario, the observed peak of
the afterglow emission occurs as the spreading jet crosses our viewing
angle.  The timescale for this peak is $\sim 100$ days for a GRB jet
with typical parameters ($E_K=10^{51}$ erg, $n=1~\rm cm^{-3}$,
$\epsilon_e=\epsilon_B=0.1$, $\theta_j=5^{\circ}$) viewed from an
angle $\theta_{\rm obs}=30^{\circ}$ \citep{snb+06}.  The subsequent
afterglow evolution is the same as that seen by an on-axis
observer, decaying steeply with $\alpha=-p\approx [-2$ to $-3]$ for
frequencies above $\nu_m$ \citep{sph99}.  The observed timescale and
evolution of the \grb\ radio flare are therefore roughly consistent
with the predictions of an off-axis relativistic jet.

However, as noted by \citet{mlc+06} and discussed in \S\ref{sec:jet},
the early and steady decay of the \grb\ broadband afterglow implies
the presence of relativistic ejecta directed along our line-of-sight.
These ejecta are also responsible for the production of the observed
prompt emission.  Attributing the strong late-time radio flare to an
energetic off-axis relativistic jet therefore implies that 
multiple relativistic ejecta components were produced in the
explosion.  Moreover, the steep rise and peak flux of the radio flare
imply sharp edges for the off-axis jet and a
kinetic energy 2 to 3 times larger than that of the on-axis ejecta.
While this scenario cannot be ruled out, we consider it unlikely given
the complicated ejecta geometry required.

\subsubsection{A Receding Jet}
\label{sec:receding}
\citet{ls04} describe a related scenario in which a strong late-time
radio flare is observed from a receding jet initially directed
anti-parallel to our line-of-sight.  In this case, early afterglow
emission is expected from the approaching jet while
emission from the receding jet is expected on a timescale
$5\times t_{\rm NR}$ due to the light travel time delay.  This
scenario is appealing in that it may explain both the early and
late-time afterglow emission observed for \grb\ within the standard
framework of engine-driven (accretion-fed compact source; \citealt{pir99})
double-sided jets.  However, as discussed in
\S\ref{sec:jet}, broadband modeling of the afterglow emission
predicts the non-relativistic transition to occur no earlier than
$0.6$ yrs -- too late to explain the radio flare at 40 days.
Moreover, the peak emission from a receding jet should be comparable
to the radio flux at $t_{\rm NR}$ and decay with a temporal index
given by the Sedov solution, $\alpha_{\rm rad}=-9/10$ \citep{ls04};
both of these predictions are inconsistent with the observations.  We
conclude that afterglow emission from a receding jet is unlikely to
produce the observed radio flare.

\subsection{Circumburst Density Variations}
\label{sec:density}

It has been argued that abrupt variations in the circumburst medium
can produce a strong rebrightening in the radio afterglows of GRBs.
Specifically, the dynamical interaction of the forward shock with a
wind-termination shock at $r\sim 1$ pc is predicted to cause a
rebrightening of the radio afterglow on a timescale of a few years
\citep{wij01,rdm+01,rgs+05}.  Here it is assumed that the blastwave is
expanding non-relativistically when it encounters the density jump,
overall consistent with the observed timescales for non-relativistic
transitions \citep{fwk00,bkf04,fsk+05}.  For comparison, the
interaction of a relativistic blastwave with an abrupt (step function)
density jump is not expected to cause strong variations in the
afterglow light-curves, where here it has been assumed that the
post-jump expansion is also relativistic \citep{ng06}.

In the case of \grb, the strong radio flare and X-ray plateau phase
occur on a timescale when the blastwave is still relativistic, and
therefore an abrupt circumburst density jump appears an unlikely
explanation.  This is supported by the fact that during the
relativistic regime, the flux at frequencies above $\nu_c$ should be
unaffected by circumburst density variations \citep{gs02}.  We
speculate, however, that a very large density jump may be able to
decelerate the blastwave to non-relativistic speeds on a very short
timescale and may therefore be able to explain the unusual late-time
afterglow evolution.  The increase in density would cause a shift in
$\nu_a$ which may explain the peculiar evolution of the spectral index in
the radio band (Figure~\ref{fig:vla_lt_curve}).  Based on our
afterglow modeling (\S\ref{sec:fit}), we estimate that the circumburst
radius of the forward shock at the onset of the radio flare was
roughly $\sim 1$ pc, roughly consistent with the radius of a
wind-termination shock \citep{glm96,clf04}.  In this scenario, we expect a
post-jump self-similar evolution consistent with the Sedov-Taylor
solution.  Since the late-time radio and X-ray data are not
sufficiently sensitive to trace the post-flare evolution, we cannot
rule out this possibility.

Finally, we investigated a scenario where the radio/X-ray flare is
produced by the dynamical interaction of the quasi-spherical,
non-relativistic SN ejecta with a CSM density enhancement. In fact,
density jumps have been invoked to explain radio modulations observed
for local (non-relativistic) SNe Ibc (e.g.~SN\,2003bg;
\citealt{sck+05}).  Adopting a simple minimum energy calculation
\citep{kfw+98} and requiring that the shock energy is equally
partitioned between magnetic fields and relativistic electrons, we
find that attributing the strong radio emission to the quasi-spherical
SN component requires that the SN ejecta is relativistic ($\Gamma \sim
10$) at the time of the radio/X-ray flare.  However, as discussed
above, strong flux variations are not expected while the blastwave is
relativistic.  Combined with the fact that the flare is at least a
factor of $10^2$ more radio luminous than any other SN Ibc ever
observed (including GRB-SN\,1998bw, \citealt{kfw+98}), we conclude
that the radio/X-ray flare can not be attributed to the
quasi-spherical SN ejecta.

\subsection{Energy Injection}
\label{sec:reverse}

An episode of energy injection may also cause a rebrightening of the
afterglow flux.  Energy injection may arise from long-lived central
engine activity or under the framework of a ``refreshed''
shock where a slow moving shell ejected during the initial burst
eventually catches up with the afterglow shock \citep{rm98,gnp03}.  The
observed ratio of the settling time to the epoch of injection, $\delta
t/t$, can distinguish between these two scenarios.  For $\delta t/t <
1$ the injection is produced by engine activity, while $\delta t/t
\gtrsim 1$ indicates a refreshed shock.  In the case of \grb, the
X-ray and radio data after 20 days imply $\delta t/t > 1$, suggesting
that the ejecta were refreshed on this timescale.

\begin{figure}
\plotone{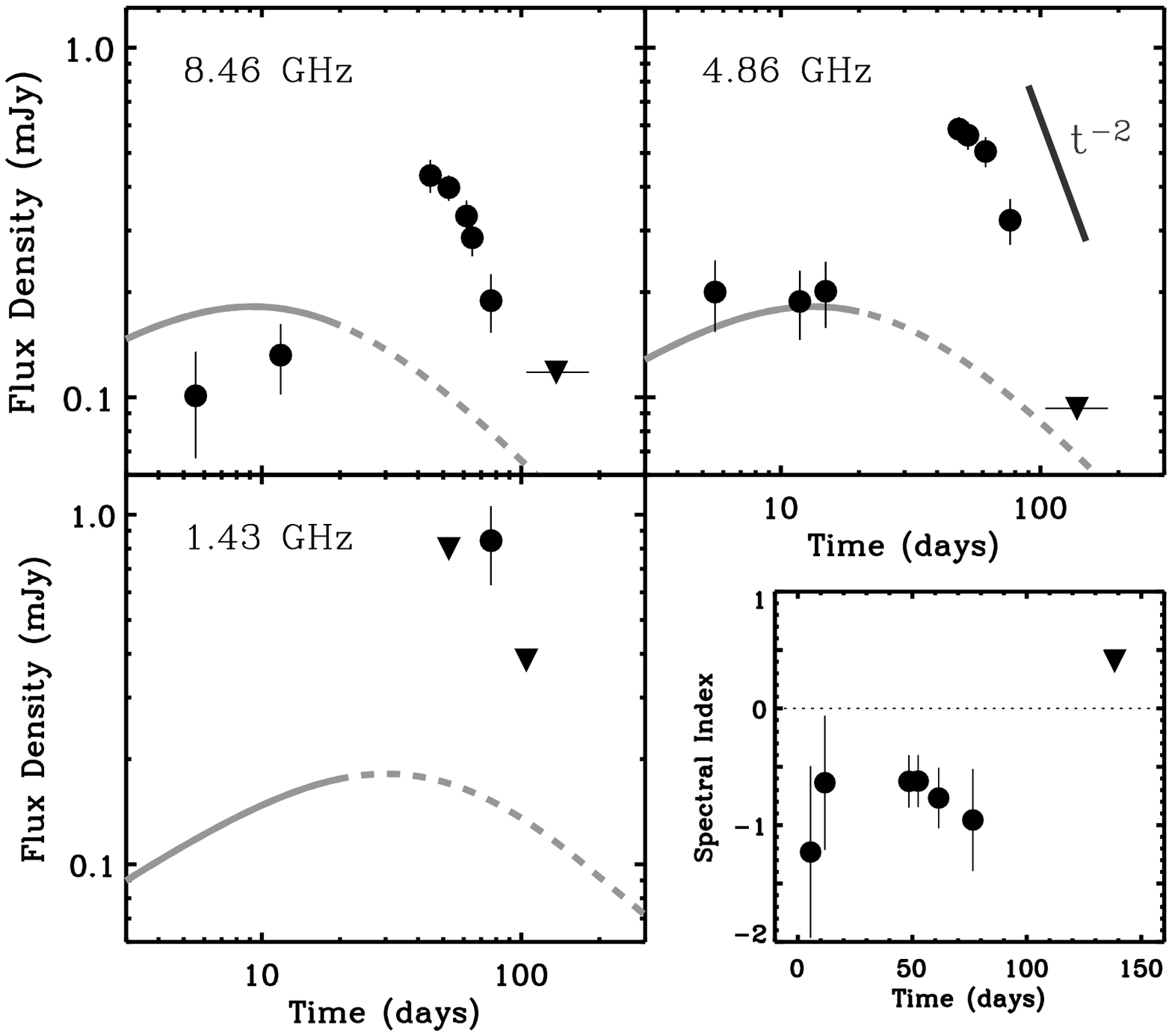}
\caption{Radio observations of \grb\ from the Very Large Array as
  listed in Table~\ref{tab:vla}.  We have summed the late-time limits
  in each frequency for clarity.  The radio evolution is characterized
  by an unusual flare at $t\sim 40$ days which rapidly fades below our
  detection limits.  This radio flare occurs roughly on the same
  timescale as the observed X-ray flattening.
  (Figure~\ref{fig:xrt_lt_curve}).  We attribute both the radio flare
  and the X-ray flattening to an episode of late-time energy injection
  (refreshing shell or off-axis ejecta) or a large circumburst density
  jump.  Our early afterglow model fit is shown (solid grey lines) and
  extrapolated to the epoch of the radio flare (dashed grey lines).
  The spectral index between 8.46 and 4.86 GHz is optically thin
  throughout our radio monitoring, including the late-time flare
  (lower right panel).}
\label{fig:vla_lt_curve}
\end{figure}

As shown in Figure~\ref{fig:vla_lt_curve}, an extrapolation of the
early radio evolution lies a factor of $\sim 3$ below the observed
flux at the onset of the radio flare.  An energy increase affects the
spectral parameters according to the following scalings: $\nu_a\propto
E_{K,{\rm iso}}^{1/5}$, $\nu_m\propto E_{K,{\rm iso}}^{1/2}$,
$F_{\nu_m}\propto E_{K,{\rm iso}}$, and $\nu_c\propto E_{K,{\rm
iso}}^{-1/2}$ \citep{gs02}.  Given that $\nu_m$ is within the radio
band on this timescale we have $F_{\nu,\rm rad}\propto E_{K,{\rm
iso}}$ and thus the radio flare corresponds to an energy injection of
a comparable factor, $\sim 3$.  For the X-ray band, $F_{\nu,X}\propto
E_{K,{\rm iso}}^{(p+2)/4}$, thus for $p\approx 2.1$ an energy
injection of a factor of $\sim 3$ corresponds to a comparable increase
in the X-ray flux.  The {\it Swift}/XRT observations suggest a
flattening on this timescale.  Here we adopt the conservative
assumption that the shock microphysics ($\epsilon_e$, $\epsilon_B$ and
$p$) do not evolve during the energy injection.

One important implication of the energy injection model is that the
post-injection asymptotic temporal decay should be the same as that
before the injection.  This prediction is consistent with the energy
injection episodes invoked for GRBs 021004 and 030329
\citep{bgj04,gnp03}.  However, in the case of \grb, the late-time
radio and X-ray data suggest a steep post-injection decay,
$\alpha\gtrsim -2$, significantly steeper than that observed
pre-injection.  Moreover, the afterglow should asymptotically approach
a flux normalization larger by a factor of $\sim 3$ in both the radio
and X-ray bands.  As shown in Figures~\ref{fig:vla_lt_curve} and
\ref{fig:xrt_lt_curve}, this appears inconsistent with the
observations which suggest that the late-time afterglow is comparable
(or fainter) than an extrapolation of the early afterglow model.  We
note, however, that the faintness of the late-time afterglow and the
steep post-injection decay may be explained if a jet break occurred on
roughly the same timescale as the energy injection.

In comparison with other late-time afterglow studies
\citep{fwk00,tmg+03,bkf04,kwp+04,fsk+05}, the radio flare and X-ray
flattening observed for \grb\ are clearly atypical for long-duration
GRBs.  We therefore attribute the observed evolution to an unusual
scenario involving either a large circumburst density jump or a
late-time injection of energy (from a slow shell or off-axis ejecta).

\subsection{An Associated Supernova}
\label{sec:sn}

The {\it HST} measurements at late-time can be used to constrain any
possible contribution from an associated supernova.  Based on previous
studies of GRB-SNe, the thermal emission from an associated supernova
is predicted to reach maximum light on a timescale of $20\times (1+z)$
days with a peak magnitude of $M_{V,\rm rest}\approx -20$ mag or
fainter (\citealt{zkh04,skp+06} and references therein).
Observationally, the emergence of a SN component produces a steepening
of the optical spectrum as the supernova nears maximum light and
dominates the afterglow emission.

An extrapolation of the broadband afterglow model to $40$
days shows that the {\it HST} data are brighter by a factor of $\sim
5$.  For comparison, the radio flare and X-ray plateau on this same
timescale represent flux density enhancements by factors of roughly
$\sim 3$ and 2, respectively.  

As discussed in \S\ref{sec:hst} we measure the spectrum of the optical
transient within the {\it ACS} bands and find $\beta_{HST}\approx
-2.8\pm 0.3$ at 37 days.  After correction for extinction (see
\S\ref{sec:optical}), the implied spectral index becomes
$\beta_{HST}\approx -1.9$.  For comparison, the spectral index between
the radio and X-ray bands on this timescale (coincident with the radio
flare and X-ray flattening) is $\beta_{\rm radX}\approx -0.56\pm 0.04$
and the indices within the bands are $\beta_{\rm rad}\approx -0.6\pm
0.2$ (Figure~\ref{fig:vla_lt_curve}) and $\beta_X\approx -1.0$
\citep{mlc+06}, respectively.  Given that the {\it HST} derived
spectral index, even after correction for extinction, is (1)
significantly steeper than the NIR/optical index observed at early
time, $\beta_{\rm NIR/opt}\approx -1.4$, (2) significantly steeper than
$\beta_{\rm radX}$ measured at a comparable epoch, and (3)
inconsistent with the range of synchrotron spectral indices predicted
for a relativistic blastwave ($\beta=[-1.5,2.5]$, \citealt{spn98}), we
conclude that the optical flux at $t\sim 40$ days is dominated by
another emission process, likely an associated SN.

Next, to determine if the late-time optical data are consistent with
the temporal evolution of a typical GRB/XRF-associated supernova, we
compare the {\it HST} flux values with synthesized SN light-curves.
We compiled {\it UBVRIJHK} observations of SN\,1998bw from the
literature \citep{gvv+98,ms99,pcd+01} and smoothed the
extinction-corrected (Galactic component of E(B-V)=0.059,
\citealt{sfd98}) light-curves.  Here we assume that the SN\,1998bw
host-galaxy extinction is negligible, consistent with the findings of
\citet{pcd+01} based on a spectroscopic analysis of the host galaxy.
We then produced $k$-corrected NIR/optical light-curves of SN\,1998bw
at the redshift of XRF\,050416a by interpolating over the photometric
spectrum and stretching the arrival time of the photons by a factor of
$(1+z)$.

Shown in Figure~\ref{fig:NIRopt_model_ltcurves2} are the synthesized
light-curves for SN\,1998bw at $z=0.6528$, summed together with the
afterglow model.  The {\it HST} data are roughly comparable with the
flux normalization and evolution of the summed model.  Therefore, the
temporal and spectral properties of the {\it HST} data suggest that
XRF\,050416a was associated with a supernova similar to SN\,1998bw.
However, we caution that the temporal coincidence of the SN peak with
the radio flare makes it difficult to estimate the relative
contributions of the SN and afterglow.

\section{Host Galaxy Properties}
\label{sec:host}

We now turn to the properties of the GRB host galaxy.  We measure the
brightness of the host galaxy in the final {\it HST} epoch to be
F775W$=23.1\pm 0.1$ mag ($I=22.7\pm 0.1$ mag).  These
values are not corrected for extinction.  At $z=0.6528$ the rest-frame
$B$-band is traced by the observed F775W band, leading to an absolute
magnitude, $M_B\approx -20.3\pm 0.1$ mag, or a luminosity $L_B\approx
0.5L^{*}$.  This host luminosity is similar to that inferred for
XRF\,030528 \citep{rsg05} and the hosts of typical GRBs
\citep{ldm+03}.  At $z=0.6528$, the measured offset of the optical
transient (\S\ref{sec:optical}) relative to the center of the host
galaxy corresponds to $140\pm 140$ pc.  This offset is a factor of $\sim 10$
smaller than the median value for long-duration GRBs \citep{bkd02}.

As shown in Figure~\ref{fig:lris_spec}, the host exhibits several
emission lines typical of star-forming galaxies.  We estimate the star
formation rate in the host galaxy from the observed fluxes of the
various emission lines.  Using the flux of the [\ion{O}{2}]$\lambda
3727$ line, $F_{\rm [OII]}\approx 9.6\times 10^{-17}$ erg cm$^{-2}$
s$^{-1}$ (Table~\ref{tab:lines}), and the conversion of \citet{ken98},
${\rm SFR}=(1.4\pm 0.4)\times 10^{-41}L_{\rm [OII]}$ M$_\odot$
yr$^{-1}$, we find a star formation rate of about $2.5\pm 0.7$
M$_\odot$ yr$^{-1}$.  From the H$\beta$ line flux, $F_{\rm
H\beta}\approx 3.7\times 10^{-17}$ erg cm$^{-2}$ s$^{-1}$, and
assuming the Case-B recombination ratio of $F_{\rm H\alpha}/F_{\rm
H\beta}=2.87$ and the conversion of \citet{ken98}, we infer a star
formation rate of ${\rm SFR}=7.9\times 10^{-42} L_{\rm H\alpha}\approx
1.5\pm 0.2$ M$_\odot$ yr$^{-1}$.  Thus, we conclude that the star
formation rate (not corrected for extinction) is roughly
$2~$M$_\odot$ yr$^{-1}$.  We note that the observed ratio of
H$\gamma$/H$\beta$ $=0.3\pm 0.1$ compared to the theoretical
value of about $0.47$ \citep{ost89} suggests a significant extinction
correction (factor of $\sim 10$) following the method of
\citep{cks94}.  We conclude that the star formation rate for the host
of XRF\,050416a is similar to those inferred for long-duration GRBs
\citep{chg04,bck+03} and at least an order of magnitude larger than
that inferred for XRF\,020903 \citep{skb+04a,bfs+06}.

The combination of the inferred star formation rate and host
luminosity provides a measure of the specific star formation rate.  We
find a value of $4$ M$_\odot$ yr$^{-1}$ $L$*/$L$ (uncorrected for
extinction), which is about a factor of two lower than the mean
specific star formation rate for the hosts of long-duration GRBs
\citep{chg04}.

Next, we use the relative strengths of the oxygen and hydrogen
emission lines to infer the ionization state and oxygen abundance.
The relevant indicators are $R_{23}\equiv {\rm log}\,(F_{\rm
[OII]}+F_{\rm [OIII]}/F_{\rm H\beta})\approx 0.74$ and $O_{32}\equiv
{\rm log}\,(F_{\rm [OIII]}/ F_{\rm [OII]})\approx 0.055$.  Using the
calibrations of \citet{mcg91} and \citet{kk04} we find that for the
upper branch the metallicity is $12+{\rm log}\,({\rm O/H})\approx 8.6$
while for the lower branch it is about 7.9; the two branches are due
to the double-valued nature of $R_{23}$ in terms of metallicity.
Thus, the host metallicity of \grb\ is $0.2-0.8$ $Z_{\odot}$, larger
than that inferred for XRF\,060218 \citep{msg+06,wsv+07} and
comparable to that for XRF\,030528 \citep{rsg05}.  Moreover, this
range is somewhat higher than the typical metallicities for GRB hosts,
some of which have metallicities that are $\sim 1/10$ solar (e.g.,
\citealt{pbc+05}).

\section{Is \grb\ a Short Burst?}
\label{sec:shb}

With a prompt emission duration of just T$_{90}\approx 2.4$ sec
\citep{sbb+06}, it is interesting to consider \grb\ as a member of the
short-hard class of gamma-ray bursts, popularly believed to result
from the coalescence of neutron stars or black holes (e.g.,
\citealt{elp+89}). Based on the bi-modal BATSE duration distribution,
bursts with $T_{90}\gtrsim 2$ seconds are assumed to belong to the
long-duration class \citep{kmf+93}, although a decomposition of the
overlapping distributions suggests that a small fraction of
short-hard bursts (SHBs) have durations longer than this cut-off.  The
distinction between short-hard and long-duration bursts is further
complicated by the detection of soft X-ray tails lasting several
seconds following SHBs 050709 and 050724 \citep{vlr+05,bcb+05}.  This
suggests that SHBs are not necessarily characterized by a pure hard
emission spectrum \citep{sak06}.  Related to this issue is the use of
spectral lags to distinguish between long and short bursts. As
discussed by \citet{nb06}, long-duration bursts typically have longer
lags that correlate with isotropic equivalent prompt gamma-ray
luminosity \citep{nmb00}.  On the other hand, SHBs have negligible (or
even negative) spectral lags.

In the case of \grb, the prompt duration places it between the long-
and short-duration classes.  The low value of $E_p$ may suggest that
it belongs to the long-duration class, however, it is becoming clear
that hardness cannot be used to reliably distinguish between the two
classes (see \citealt{udi07}).
Similarly, while the negative lag inferred for \grb\ may suggest a SHB
classification, examples do exist of long-duration BATSE bursts with
negative lags \citep{nb06}.  

Afterglow modeling may provide additional clues.  The range of values
we infer for the \grb\ beaming-corrected energies are overall
consistent with those of long-duration GRBs including the subclass of
sub-energetic bursts.  At the same time, they are roughly consistent
with the values inferred for SHBs
\citep{ffp+05,bpc+05,sbk+06}.  However, the low circumburst density,
$n\sim 10^{-3}$, is a factor of 10 to $10^4$ smaller than the typical
values inferred for long-duration GRBs (e.g.
\citealt{pk02,yhs+03,clf04}) but comparable to those of SHBs
\citep{ffp+05,bpc+05,bpp+06,bgc+06,pan06,sbk+06}.  The observed radio
flare and X-ray flattening at 20 days are atypical for both long- and
short- duration GRBs and therefore cannot be used to classify this
event.  However, it is interesting to note that large variations in
the circumburst density are more naturally explained in the context of
a massive stellar progenitor with interacting stellar winds
\citep{glm96,rdm+01}.  

Next is a discussion of the \grb\ host galaxy since long- and short-
bursts may also be distinguished by their environments (see
\citealt{ber06} for a recent review).  As discussed in
\S\ref{sec:host}, the host is a star-forming galaxy with an inferred
SFR and metallicity comparable to those of long-duration GRBs.
Moreover, \grb\ is located near the center and brightest part of its
host galaxy.  It is thus consistent with the locations of
long-duration bursts with respect to their hosts \citep{bkd02,fls+06}.
For comparison, SHBs are typically localized to low SFR hosts with
significant old stellar populations at radial offsets up to a factor
of 10 larger than those of typical long-duration bursts
\citep{bpc+05,ffp+05,bpc+06,sbk+06}.

Finally, the discovery of an associated Type Ic supernova is perhaps
one of the best methods to distinguish between long- and short-bursts.
The discovery of several long-duration bursts and X-ray flashes at
$z\lesssim 0.3$ in the last few years has firmly established that GRBs
and XRFs are accompanied by supernovae of Type Ic
\citep{gvv+98,smg+03,mtc+04,skf+05,pmm+06}.  A study of the luminosity
distribution for GRB/XRF-SNe reveals a significant dispersion,
implying a spread of (at least) an order of magnitude in peak optical
luminosity \citep{zkh04,skp+06}.  At the same time, deep imaging of
SHBs has constrained any associated SN emission to be up to $\sim 100$
times less luminous as SN\,1998bw \citep{ffp+05,sbk+06}.  In the case
of \grb, the temporal and spectral properties of the {\it HST} data
suggest that \grb\ was accompanied by a supernova with a peak
luminosity roughly similar to SN\,1998bw.  We stress, however, that
the temporal coincidence of the radio flare with the SN peak
complicates any study of the SN properties.

Based on the associated SN and large-scale environmental properties,
we conclude that \grb\ is a member of the long-duration class of gamma-ray
bursts.  This event highlights the difficulty in classifying bursts
based on their prompt emission properties alone.

\section{Discussion and Conclusions}
\label{sec:disc}

We present extensive broad-band NIR/optical and radio data for the
afterglow of the X-ray flash 050416a and show that it is localized to
a star-forming ($\gtrsim 2~$M$_\odot$ yr$^{-1}$) host galaxy at
$z=0.6528$.  Along with XRFs 020903, 030723 and 060218, this burst is
one of the best studied X-ray flashes to date.  Moreover, \grb\ is
only the third XRF with a spectroscopic redshift for which a broadband
afterglow study has been performed and the physical parameters have
been constrained.  The isotropic-equivalent prompt and kinetic energy
releases are $E_{\gamma,\rm iso}\approx 1.2\times 10^{51}$
\citep{sbb+06} and $E_{K,\rm iso}\approx 10^{52}$ erg respectively.
These values are a factor of $10^2$ times larger than those of
XRF\,020903 \citep{slg+04,skb+04a} and up to $10^4$ times larger than
those inferred for sub-energetic GRBs 980425, 031203 and XRF\,060218
\citep{paa+00,kfw+98,sls04,skb+04b,cmb+06,skn+06}.

Adopting the results of our standard synchrotron model and $t_j\gtrsim
20$ days we constrain the collimation of the ejecta to
$\theta_j\gtrsim (6.9-9.7)^{\circ}$, slightly larger than the median
value of $5^{\circ}$ for typical GRBs (\citealt{bfk03,ggl04} and
references therein).  This indicates that the beaming-corrected energy
release is $E_K\approx (9.8\times 10^{49} - 1.4\times 10^{52}) $ erg
and $E_{\gamma} \approx (8.6\times 10^{48} - 1.2\times 10^{51}$) erg,
implying a total relativistic energy yield of $E_{\rm tot}\approx
(1.1\times 10^{50} - 1.5\times 10^{52})$ erg, which straddles the
median value for cosmological GRBs, $E_{\rm tot}\approx 2\times
10^{51}$ erg \citep{bkp+03}.  However, the efficiency in converting
the energy in the ejecta into $\gamma$-rays is $\eta_{\gamma}\equiv
E_{\gamma}/(E_K+E_{\gamma})\approx (0.04-0.08)$, significantly lower
than the typical values for GRBs \citep{pk02,yhs+03} and comparable to
that inferred for XRF\,020903 ($\eta_{\gamma}\approx 0.03$,
\citealt{skb+04a}). This strengthens the idea that XRFs and GRBs are
distinguished by their ability to couple significant energy to
highly-relativistic material \citep{skb+04a,zwh04}.

In addition to the burst energetics, several key results emerge from
our broadband analysis of \grb.  First is the detection of a bright,
late-time radio flare accompanied by an observed flattening in the
X-ray bands which we attribute to a large circumburst density
enhancement or episode of energy injection (refreshing shell or
off-axis ejecta) at $t\sim 20$ days.  In the context of a density
jump, it is interesting to note that the inferred pre-jump circumburst
density is several orders of magnitude lower than the values typically
inferred for long-duration GRBs (see \citealt{snb+06} for a recent
compilation).  Moreover, radio observations of local
(optically-selected) core-collapse SNe show similar flux modulations
(factor of 2 to 3) attributed to abrupt variations in the
circumstellar medium \citep{wvd+91,rss+04,sck+05}.

Next, the temporal and spectral evolution of the optical afterglow
suggests the contribution from a supernova component with peak
luminosity and light-curve shape comparable to SN\,1998bw.  Given the
temporal coincidence of the radio flare and X-ray flattening with the
SN peak, \grb\ highlights the need for full spectral coverage in
late-time GRB/XRF-SN searches.  This is illustrated by the optical
rebrightening observed for XRF\,030723 at $\sim 15$ days and
interpretted as a thermal SN component \citep{fsh+04} while the X-ray
and radio data show similar rebrightenings at late-time suggesting a
CSM density jump or energy injection (\citealt{bss+05}, Soderberg {\it
et al.} in prep).

With the addition of \grb, there are three XRFs (020903, 060218 and
now 050416a) with spectroscopic redshifts observed in association with
supernovae with peak luminosities varying by up to a factor of 10
compared to SN\,1998bw \citep{skf+05,bfs+06,pmm+06}.  However, deep
{\it HST} observations of XRF\,040701 at $z=0.21$ revealed that any
associated SN was at least a factor of 10 (and likely $\gtrsim 100$)
times fainter than SN\,1998bw \citep{skf+05}.  It is therefore clear
that most XRFs produce Nickel-rich supernova explosions, but that
there is a significant dispersion in the peak luminosities of
XRF-associated SNe.  A similar result is found for GRB-associated SNe
\citep{zkh04,skp+06} and further strengthens the idea that GRBs and
XRFs are intimately related.

Finally, we address \grb\ within the framework of off-axis and dirty
fireball models for XRFs.  As discussed in \S\ref{sec:jet}, the
evolution of the early afterglow indicates that the ejecta are being
viewed along the collimation axis.  Similarly, on-axis viewing angles
are inferred from afterglow studies of XRFs 020903 \citep{skb+04a},
050215B \citep{lot+06}, 050406 \citep{rmb+06}, and 060218
\citep{cmb+06,skn+06}.  Unification models in which X-ray flashes are
understood as typical GRBs viewed away from the burst collimation axis
(e.g. \citealt{gpk+02,yyn+03}) are therefore inconsistent with the
observations for these five events.  To date, such models are only
consistent with the observations of one X-ray flash: XRF\,030723
\citep{fsh+04,bss+05,grp05}.

While the lack of high energy photons prevents a direct constraint on
the Lorentz factor of the burst, our detailed afterglow modeling
allows us to place a lower limit on the initial bulk Lorentz factor,
$\Gamma_0\gtrsim 110$, by extrapolating $\Gamma \propto t^{-3/8}$ back
to the first XRT observation at $\sim 100$ sec.  This value is
somewhat lower than the typical values inferred for cosmological GRBs
\citep{pk02} and consistent with the upper range of values predicted
for a dirty fireball.  Given that some dirty fireball models predict
XRFs to have wider jets than typical GRBs (e.g., \citealt{zwm03}), it
is interesting to note that there are no X-ray flashes for which an
achromatic jet break has been observed. This may suggest that their
ejecta are not strongly collimated.  In fact, detailed radio
observations of XRFs 020903 and 060218 spanning $\gtrsim 100$ days
imply a quasi-spherical ejecta geometry in both cases
\citep{skb+04a,skn+06}.

While the basic properties of X-ray flashes (redshifts, hosts,
isotropic $\gamma$-ray energies) are now available for several events,
it is clear that broadband afterglow observations are required for a
complete understanding of the burst properties.  With the addition of
\grb\ to the existing sample of only a few well-studied events, we
continue to work towards a systematic comparison of the global
characteristics of XRFs and GRBs.  While limited, the current sample
suggests that the two classes share similar ejecta energies,
associated SNe, and viewing angles.  Further studies of XRF afterglows
will be used to confirm whether the two classes differ in their
collimation angles.

The authors thank Re'em Sari and Don Lamb for helpful discussions.  As
always, the authors thank Jochen Greiner for maintaining his GRB page.
A.M.S. and S.B.C. are supported by the NASA Graduate Student Research
Program.  E.B. and A.G. acknowledge support by NASA through Hubble
Fellowship grants awarded by STScI, which is operated by AURA, Inc.,
for NASA.  GRB research at Caltech is supported through NASA.

\bibliographystyle{apj1b}

\begin{thebibliography}{}

\bibitem[{Alard}(2000)]{a00}
{Alard}, C. 2000, \aaps, 144, 363.

\bibitem[{Amati}(2006)]{ama06}
{Amati}, L. 2006, \mnras, 2006, 372, 233.

\bibitem[{Band} {\it et al.}\ (1993)]{bmf+93}
{Band}, D. {\it et al.}\  1993, \apj, 413, 281.

\bibitem[{Barthelmy} {\it et al.}\ (2005)]{bcb+05}
{Barthelmy}, S.~D. {\it et al.}\  2005, \nat, 438, 994.

\bibitem[{Berger}(2006)]{ber06}
{Berger}, E. 2006, in { AIP Conf. Proc. 838: Gamma-Ray Bursts in the Swift
  Era}, ed.\ S.~S. {Holt}, N. {Gehrels}, and J.~A. {Nousek}, 33.

\bibitem[{Berger} {\it et al.}\ (2003a)]{bck+03}
{Berger}, E., {Cowie}, L.~L., {Kulkarni}, S.~R., {Frail}, D.~A., {Aussel}, H.,
  and {Barger}, A.~J. 2003a, \apj, 588, 99.

\bibitem[{Berger}, {Kulkarni} \& {Frail}(2003)]{bkf03}
{Berger}, E., {Kulkarni}, S.~R., and {Frail}, D.~A. 2003, \apj, 590, 379.

\bibitem[{Berger}, {Kulkarni} \& {Frail}(2004)]{bkf04}
{Berger}, E., {Kulkarni}, S.~R., and {Frail}, D.~A. 2004, \apj, 612, 966.

\bibitem[{Berger} {\it et al.}\ (2003b)]{bkp+03}
{Berger}, E. {\it et al.}\  2003b, \nat, 426, 154.

\bibitem[{Berger} {\it et al.}\ (2005)]{bpc+05}
{Berger}, E. {\it et al.}\  2005, \nat, 438, 988.

\bibitem[{Bersier} {\it et al.}\ (2006)]{bfs+06}
{Bersier}, D. {\it et al.}\  2006, \apj, 643, 284.

\bibitem[{Bj{\"o}rnsson}, {Gudmundsson} \& {J{\'o}hannesson}(2004)]{bgj04}
{Bj{\"o}rnsson}, G., {Gudmundsson}, E.~H., and {J{\'o}hannesson}, G. 2004,
  \apjl, 615, L77.

\bibitem[{Bloom} {\it et al.}\ (2003)]{bfv+03}
{Bloom}, J.~S., {Fox}, D., {van Dokkum}, P.~G., {Kulkarni}, S.~R., {Berger},
  E., {Djorgovski}, S.~G., and {Frail}, D.~A. 2003, \apj, 599, 957.

\bibitem[{Bloom}, {Frail} \& {Kulkarni}(2003)]{bfk03}
{Bloom}, J.~S., {Frail}, D.~A., and {Kulkarni}, S.~R. 2003, \apj, 594, 674.

\bibitem[{Bloom}, {Kulkarni} \& {Djorgovski}(2002)]{bkd02}
{Bloom}, J.~S., {Kulkarni}, S.~R., and {Djorgovski}, S.~G. 2002, \aj, 123,
  1111.

\bibitem[{Bloom} {\it et al.}\ (2006a)]{bpc+06}
{Bloom}, J.~S. {\it et al.}\  2006a, astro-ph/0607223

\bibitem[{Bloom} {\it et al.}\ (2006b)]{bpp+06}
{Bloom}, J.~S. {\it et al.}\  2006b, \apj, 638, 354.

\bibitem[{Burrows} {\it et al.}\ (2006)]{bgc+06}
{Burrows}, D.~N. {\it et al.}\  2006, \apj, 653, 468.

\bibitem[{Butler} {\it et al.}\ (2005)]{bss+05}
{Butler}, N.~R. {\it et al.}\ 2005, \apj, 621, 884.

\bibitem[{Calzetti}, {Kinney} \& {Storchi-Bergmann}(1994)]{cks94}
{Calzetti}, D., {Kinney}, A.~L., and {Storchi-Bergmann}, T. 1994, \apj, 429,
  582.

\bibitem[{Campana} {\it et al.}\ (2006)]{cmb+06}
{Campana}, S. {\it et al.}\  2006, \nat, 442, 1008.

\bibitem[{Chevalier}, {Li} \& {Fransson}(2004)]{clf04}
{Chevalier}, R.~A., {Li}, Z.-Y., and {Fransson}, C. 2004, \apj, 606, 369.

\bibitem[{Christensen}, {Hjorth} \& {Gorosabel}(2004)]{chg04}
{Christensen}, L., {Hjorth}, J., and {Gorosabel}, J. 2004, \aap, 425, 913.

\bibitem[{Dermer}, {Chiang} \& {B{\"o}ttcher}(1999)]{dcb99}
{Dermer}, C.~D., {Chiang}, J., and {B{\"o}ttcher}, M. 1999, \apj, 513, 656.

\bibitem[{Eichler} {\it et al.}\ (1989)]{elp+89}
{Eichler}, D., {Livio}, M., {Piran}, T., and {Schramm}, D.~N. 1989, \nat, 340,
  126.

\bibitem[{Fox} {\it et al.}\ (2005)]{ffp+05}
{Fox}, D.~B. {\it et al.}\  2005, \nat, 437, 845.

\bibitem[{Frail} {\it et al.}\ (2003)]{fkb+03}
{Frail}, D.~A., {Kulkarni}, S.~R., {Berger}, E., and {Wieringa}, M.~H. 2003,
  \aj, 125, 2299.

\bibitem[{Frail} {\it et al.}\ (2001)]{fks+01}
{Frail}, D.~A. {\it et al.}\  2001, \apjl, 562, L55.

\bibitem[{Frail} {\it et al.}\ (2005)]{fsk+05}
{Frail}, D.~A., {Soderberg}, A.~M., {Kulkarni}, S.~R., {Berger}, E., {Yost},
  S., {Fox}, D.~W., and {Harrison}, F.~A. 2005, \apj, 619, 994.

\bibitem[{Frail}, {Waxman} \& {Kulkarni}(2000)]{fwk00}
{Frail}, D.~A., {Waxman}, E., and {Kulkarni}, S.~R. 2000, \apj, 537, 191.

\bibitem[{Freedman} \& {Waxman}(2001)]{fw01}
{Freedman}, D.~L. and {Waxman}, E. 2001, \apj, 547, 922.

\bibitem[{Fruchter} \& {Hook}(2002)]{fh02}
{Fruchter}, A.~S. and {Hook}, R.~N. 2002, \pasp, 114, 144.

\bibitem[{Fruchter} {\it et al.}\ (2006)]{fls+06}
{Fruchter}, A.~S. {\it et al.}\  2006, \nat, 441, 463.

\bibitem[{Fynbo} {\it et al.}\ (2004)]{fsh+04}
{Fynbo}, J.~P.~U. {\it et al.}\ 2004, \apj, 609, 962.

\bibitem[{Galama} {\it et al.}\ (1998)]{gvv+98}
{Galama}, T.~J. {\it et al.}\  1998, \nat, 395, 670.

\bibitem[{Galama} \& {Wijers}(2001)]{gw01}
{Galama}, T.~J. and {Wijers}, R.~A.~M.~J. 2001, \apjl, 549, L209.

\bibitem[{Garcia-Segura}, {Langer} \& {Mac Low}(1996)]{glm96}
{Garcia-Segura}, G., {Langer}, N., and {Mac Low}, M.-M. 1996, \aap, 316, 133.

\bibitem[{Ghirlanda}, {Ghisellini} \& {Lazzati}(2004)]{ggl04}
{Ghirlanda}, G., {Ghisellini}, G., and {Lazzati}, D. 2004, \apj, 616, 331.

\bibitem[{Granot}, {Nakar} \& {Piran}(2003)]{gnp03}
{Granot}, J., {Nakar}, E., and {Piran}, T. 2003, \nat, 426, 138.

\bibitem[{Granot} {\it et al.}\ (2002)]{gpk+02}
{Granot}, J., {Panaitescu}, A., {Kumar}, P., and {Woosley}, S.~E. 2002, \apjl,
  570, L61.

\bibitem[{Granot} \& {Sari}(2002)]{gs02}
{Granot}, J. and {Sari}, R. 2002, \apj, 568, 820.

\bibitem[{Granot}, {Ramirez-Ruiz} \& {Perna}\ (2005)]{grp05}
{Granot}, J. and {Ramirez-Ruiz}, E. and {Perna}, R. 2005, \apj, 630, 1003.

\bibitem[{Heise} {\it et al.}\ (2001)]{hik+01}
{Heise}, J., {in't Zand}, J., {Kippen}, R.~M., and {Woods}, P.~M. 2001, in {
  Gamma-ray Bursts in the Afterglow Era}, ed.\ E. {Costa}, F. {Frontera}, and
  J. {Hjorth}, 16.

\bibitem[{Henden}(2005)]{hen05}
{Henden}, A. 2005, GRB Coordinates Network, 3454, 1.

\bibitem[{Holland} {\it et al.}\ (2006)]{hbg+06}
{Holland}, S.~T. {\it et al.}\  2006, ArXiv Astrophysics e-prints.
\newblock astro-ph/0604316.

\bibitem[{Jakobsson} {\it et al.}\ (2004)]{jhf+04}
{Jakobsson}, P. {\it et al.}\  2004, \aap, 427, 785.

\bibitem[{Kennicutt}(1998)]{ken98}
{Kennicutt}, R.~C. 1998, \araa, 36, 189.

\bibitem[{Kobulnicky} \& {Kewley}(2004)]{kk04}
{Kobulnicky}, H.~A. and {Kewley}, L.~J. 2004, \apj, 617, 240.

\bibitem[{Kouveliotou} {\it et al.}\ (1993)]{kmf+93}
{Kouveliotou}, C., {Meegan}, C.~A., {Fishman}, G.~J., {Bhat}, N.~P., {Briggs},
  M.~S., {Koshut}, T.~M., {Paciesas}, W.~S., and {Pendleton}, G.~N. 1993,
  \apjl, 413, L101.

\bibitem[{Kouveliotou} {\it et al.}\ (2004)]{kwp+04}
{Kouveliotou}, C. {\it et al.}\  2004, \apj, 608, 872.

\bibitem[{Kulkarni} {\it et al.}\ (1999)]{kfs+99}
{Kulkarni}, S.~R. {\it et al.}\  1999, \apjl, 522, L97.

\bibitem[{Kulkarni} {\it et al.}\ (1998)]{kfw+98}
{Kulkarni}, S.~R. {\it et al.}\  1998, \nat, 395, 663.

\bibitem[{Le Floc'h} {\it et al.}\ (2003)]{ldm+03}
{Le Floc'h}, E. {\it et al.}\  2003, \aap, 400, 499.

\bibitem[{Levan} {\it et al.}\ (2006)]{lot+06}
{Levan}, A.~J. {\it et al.}\  2006, \apj, 648, 1132.

\bibitem[{Li} {\it et al.}\ (2005)]{lcj+05}
{Li}, W., {Chornock}, R., {Jha}, S., and {Filippenko}, A.~V. 2005, GRB
  Coordinates Network, 3270, 1.

\bibitem[{Li} \& {Song}(2004)]{ls04}
{Li}, Z. and {Song}, L.~M. 2004, \apjl, 614, L17.

\bibitem[{Lithwick} \& {Sari}(2001)]{ls01}
{Lithwick}, Y. and {Sari}, R. 2001, \apj, 555, 540.

\bibitem[{Livio} \& {Waxman}(2000)]{lw00}
{Livio}, M. and {Waxman}, E. 2000, \apj, 538, 187.

\bibitem[{Malesani} {\it et al.}\ (2004)]{mtc+04}
{Malesani}, D. {\it et al.}\  2004, \apjl, 609, L5.

\bibitem[{Mangano} {\it et al.}\ (2006)]{mlc+06}
{Mangano}, V. {\it et al.}\  2006, ArXiv Astrophysics e-prints.
\newblock astro-ph/0603738.

\bibitem[{McGaugh}(1991)]{mcg91}
{McGaugh}, S.~S. 1991, \apj, 380, 140.

\bibitem[{McKenzie} \& {Schaefer}(1999)]{ms99}
{McKenzie}, E.~H. and {Schaefer}, B.~E. 1999, \pasp, 111, 964.

\bibitem[{Mirabal} {\it et al.}\ (2006)]{mha+06}
{Mirabal}, N., {Halpern}, J.~P., {An}, D., {Thorstensen}, J.~R., and
  {Terndrup}, D.~M. 2006, \apjl, 643, L99.

\bibitem[{Modjaz} {\it et al.}\ (2006)]{msg+06}
{Modjaz}, M. {\it et al.}\  2006, \apjl, 645, L21.

\bibitem[{Nakar} \& {Granot}(2006)]{ng06}
{Nakar}, E. and {Granot}, J. 2006, ArXiv Astrophysics e-prints.
\newblock {astro-ph/0606011}.

\bibitem[{Nakar}(2007)]{udi07}
{Nakar}, E. 2007, Physics Reports (in press).
\newblock {astro-ph/0701748}.

\bibitem[{Norris} \& {Bonnell}(2006)]{nb06}
{Norris}, J.~P. and {Bonnell}, J.~T. 2006, \apj, 643, 266.

\bibitem[{Norris}, {Marani} \& {Bonnell}(2000)]{nmb00}
{Norris}, J.~P., {Marani}, G.~F., and {Bonnell}, J.~T. 2000, \apj, 534, 248.

\bibitem[{Nousek} {\it et al.}\ (2006)]{nkg+06}
{Nousek}, J.~A. {\it et al.}\  2006, \apj, 642, 389.

\bibitem[{O'Brien} {\it et al.}\ (2006)]{owo+06}
{O'Brien}, P.~T. {\it et al.}\  2006, \apj, 647, 1213.

\bibitem[{Osterbrock}(1989)]{ost89}
{Osterbrock}, D.~E. 1989, { {Astrophysics of gaseous nebulae and active
  galactic nuclei}}, ({University of Minnesota, et al.~Mill Valley, CA}:
  Research supported by the University of California, John Simon Guggenheim
  Memorial Foundation, University Science Books).

\bibitem[{Paczynski}(2001)]{pac01}
{Paczynski}, B. 2001, Acta Astronomica, 51, 1.

\bibitem[{Panaitescu}(2006)]{pan06}
{Panaitescu}, A. 2006, \mnras, 367, L42.

\bibitem[{Panaitescu} \& {Kumar}(2002)]{pk02}
{Panaitescu}, A. and {Kumar}, P. 2002, \apj, 571, 779.

\bibitem[{Patat} {\it et al.}\ (2001)]{pcd+01}
{Patat}, F. {\it et al.}\  2001, \apj, 555, 900.

\bibitem[{Pei}(1992)]{pei92}
{Pei}, Y.~C. 1992, \apj, 395, 130.

\bibitem[{Pian} {\it et al.}\ (2000)]{paa+00}
{Pian}, E. {\it et al.}\  2000, \apj, 536, 778.

\bibitem[{Pian} {\it et al.}\ (2006)]{pmm+06}
{Pian}, E. {\it et al.}\  2006, \nat, 442, 1011.

\bibitem[{Piran}(1999)]{pir99}
{Piran}, T. 1999, \physrep, 314, 575.

\bibitem[{Predehl} \& {Schmitt}(1995)]{ps95}
{Predehl}, P. and {Schmitt}, J.~H.~M.~M. 1995, \aap, 293, 889.

\bibitem[{Price} {\it et al.}\ (2005)]{pmc+05}
{Price}, P.~A., {Minezaki}, T., {Cowie}, L., and {Yoshii}, Y. 2005, GRB
  Coordinates Network, 3312, 1.

\bibitem[{Prochaska} {\it et al.}\ (2006)]{pbc+05}
{Prochaska}, J.~X. {\it et al.}\  2006, \apj, 642, 989.

\bibitem[{Qiu} {\it et al.}\ (2005)]{qll+05}
{Qiu}, Y., {Lu}, C.~L., {Lou}, Y.~Q., {Huang}, K.~Y., and {Urata}, Y. 2005, GRB
  Coordinates Network, 3286, 1.

\bibitem[{Ramirez-Ruiz} {\it et al.}\ (2001)]{rdm+01}
{Ramirez-Ruiz}, E., {Dray}, L.~M., {Madau}, P., and {Tout}, C.~A. 2001, \mnras,
  327, 829.

\bibitem[{Ramirez-Ruiz} {\it et al.}\ (2005)]{rgs+05}
{Ramirez-Ruiz}, E., {Garc{\'{\i}}a-Segura}, G., {Salmonson}, J.~D., and
  {P{\'e}rez-Rend{\'o}n}, B. 2005, \apj, 631, 435.

\bibitem[{Rau}, {Salvato} \& {Greiner}(2005)]{rsg05}
{Rau}, A., {Salvato}, M., and {Greiner}, J. 2005, \aap, 444, 425.

\bibitem[{Rees} \& {Meszaros}(1998)]{rm98}
{Rees}, M.~J. and {Meszaros}, P. 1998, \apjl, 496, L1+.

\bibitem[{Romano} {\it et al.}\ (2006)]{rmb+06}
{Romano}, P. {\it et al.}\  2006, \aap, 450, 59.

\bibitem[{Ryder} {\it et al.}\ (2004)]{rss+04}
{Ryder}, S.~D., {Sadler}, E.~M., {Subrahmanyan}, R., {Weiler}, K.~W.,
  {Panagia}, N., and {Stockdale}, C. 2004, \mnras, 349, 1093.

\bibitem[{Sakamoto}(2006)]{sak06}
{Sakamoto}, T. 2006, in { AIP Conf. Proc. 838: Gamma-Ray Bursts in the Swift
  Era}, ed.\ S.~S. {Holt}, N. {Gehrels}, and J.~A. {Nousek}, 43.

\bibitem[{Sakamoto} {\it et al.}\ (2006)]{sbb+06}
{Sakamoto}, T. {\it et al.}\  2006, \apjl, 636, L73.

\bibitem[{Sakamoto} {\it et al.}\ (2004)]{slg+04}
{Sakamoto}, T. {\it et al.}\  2004, \apj, 602, 875.

\bibitem[{Sakamoto} {\it et al.}\ (2005)]{slk+05}
{Sakamoto}, T. {\it et al.}\  2005, \apj, 629, 311.

\bibitem[{Sari} \& {Piran}(1999)]{sp99}
{Sari}, R. and {Piran}, T. 1999, \apjl, 517, L109.

\bibitem[{Sari}, {Piran} \& {Halpern}(1999)]{sph99}
{Sari}, R., {Piran}, T., and {Halpern}, J.~P. 1999, \apjl, 519, L17.

\bibitem[{Sari}, {Piran} \& {Narayan}(1998)]{spn98}
{Sari}, R., {Piran}, T., and {Narayan}, R. 1998, \apjl, 497, L17+.

\bibitem[{Sazonov}, {Lutovinov} \& {Sunyaev}(2004)]{sls04}
{Sazonov}, S.~Y., {Lutovinov}, A.~A., and {Sunyaev}, R.~A. 2004, \nat, 430,
  646.

\bibitem[{Schady} {\it et al.}\ (2006)]{smo+06}
{Schady}, P. {\it et al.}\  2006, \apj, 643, 276.

\bibitem[{Schlegel}, {Finkbeiner} \& {Davis}(1998)]{sfd98}
{Schlegel}, D.~J., {Finkbeiner}, D.~P., and {Davis}, M. 1998, \apj, 500, 525.

\bibitem[{Sirianni} {\it et al.}\ (2005)]{sjb+05}
{Sirianni}, M. {\it et al.}\  2005, \pasp, 117, 1049.

\bibitem[{Smith} {\it et al.}\ (2002)]{stk+02}
{Smith}, J.~A. {\it et al.}\  2002, \aj, 123, 2121.

\bibitem[{Soderberg}(2006)]{sod06}
{Soderberg}, A.~M. 2006, in { AIP Conf. Proc. 838: Gamma-Ray Bursts in the
  Swift Era}, ed.\ S.~S. {Holt}, N. {Gehrels}, and J.~A. {Nousek}, 380.

\bibitem[{Soderberg} {\it et al.}\ (2006a)]{sbk+06}
{Soderberg}, A.~M. {\it et al.}\  2006a, \apj, 650, 261.

\bibitem[{Soderberg} {\it et al.}\ (2006b)]{sck+05}
{Soderberg}, A.~M., {Chevalier}, R.~A., {Kulkarni}, S.~R., and {Frail}, D.~A.
  2006b, \apj, 651, 1005.

\bibitem[{Soderberg} {\it et al.}\ (2004a)]{skb+04a}
{Soderberg}, A.~M. {\it et al.}\  2004a, \apj, 606, 994.

\bibitem[{Soderberg} {\it et al.}\ (2004b)]{skb+04b}
{Soderberg}, A.~M. {\it et al.}\  2004b, \nat, 430, 648.

\bibitem[{Soderberg} {\it et al.}\ (2005b)]{skf+05}
{Soderberg}, A.~M. {\it et al.}\  2005b, \apj, 627, 877.

\bibitem[{Soderberg} {\it et al.}\ (2006c)]{skn+06}
{Soderberg}, A.~M. {\it et al.}\  2006c, \nat, 442, 1014.

\bibitem[{Soderberg} {\it et al.}\ (2006d)]{skp+06}
{Soderberg}, A.~M. {\it et al.}\  2006d, \apj, 636, 391.

\bibitem[{Soderberg} {\it et al.}\ (2006e)]{snb+06}
{Soderberg}, A.~M., {Nakar}, E., {Berger}, E., and {Kulkarni}, S.~R. 2006e,
  \apj, 638, 930.

\bibitem[{Stanek} {\it et al.}\ (2003)]{smg+03}
{Stanek}, K.~Z. {\it et al.}\  2003, \apjl, 591, L17.

\bibitem[{Tiengo} {\it et al.}\ (2003)]{tmg+03}
{Tiengo}, A., {Mereghetti}, S., {Ghisellini}, G., {Rossi}, E., {Ghirlanda}, G.,
  and {Schartel}, N. 2003, \aap, 409, 983.

\bibitem[{Torii}(2005)]{tor05}
{Torii}, K. 2005, GRB Coordinates Network, 3272, 1.

\bibitem[{Villasenor} {\it et al.}\ (2005)]{vlr+05}
{Villasenor}, J.~S. {\it et al.}\  2005, \nat, 437, 855.

\bibitem[{Waxman}(2004)]{wax04}
{Waxman}, E. 2004, \apj, 602, 886.

\bibitem[{Weiler} {\it et al.}\ (1991)]{wvd+91}
{Weiler}, K.~W., {van Dyk}, S.~D., {Discenna}, J.~L., {Panagia}, N., and
  {Sramek}, R.~A. 1991, \apj, 380, 161.

\bibitem[{Wiersema} {\it et al.}\ (2007)]{wsv+07}
{Wiersema}, K. {\it et al.}\ 2007, \aap, in press (astro-ph/0701034).

\bibitem[{Wijers}(2001)]{wij01}
{Wijers}, R.~A.~M.~J. 2001, in { Gamma-ray Bursts in the Afterglow Era}, ed.\
  E. {Costa}, F. {Frontera}, and J. {Hjorth}, 306.

\bibitem[{Yamazaki}, {Yonetoku} \& {Nakamura}(2003)]{yyn+03}
{Yamazaki}, R., {Yonetoku}, D., and {Nakamura}, T. 2003, \apjl, 594, L79.

\bibitem[{Yanagisawa}, {Toda} \& {Kawai}(2005)]{ytk05}
{Yanagisawa}, K., {Toda}, H., and {Kawai}, N. 2005, GRB Coordinates Network,
  3287, 1.

\bibitem[{Yost} {\it et al.}\ (2003)]{yhs+03}
{Yost}, S.~A., {Harrison}, F.~A., {Sari}, R., and {Frail}, D.~A. 2003, \apj,
  597, 459.

\bibitem[{Zeh}, {Klose} \& {Hartmann}(2004)]{zkh04}
{Zeh}, A., {Klose}, S., and {Hartmann}, D.~H. 2004, \apj, 609, 952.

\bibitem[{Zel'dovich} \& {Raizer}(2002)]{snt}
{Zel'dovich}, Y.~B. and {Raizer}, Y.~P. 2002, { {Physics of Shock Waves and
  High Temperature Hydrodynamic Phenomena}}, ({Mineola, NY}: Dover).

\bibitem[{Zhang}, {Woosley} \& {Heger}(2004)]{zwh04}
{Zhang}, W., {Woosley}, S.~E., and {Heger}, A. 2004, \apj, 608, 365.

\bibitem[{Zhang}, {Woosley} \& {MacFadyen}(2003)]{zwm03}
{Zhang}, W., {Woosley}, S.~E., and {MacFadyen}, A.~I. 2003, \apj, 586, 356.

\end{thebibliography}

\clearpage

\begin{deluxetable}{lrccr}
\tablecaption{Ground-based Optical and NIR Observations of XRF\,050416a}
\tablewidth{0pt} \tablehead{
\colhead{Date Obs} & \colhead{$\Delta t$\tablenotemark{a}} & \colhead{Telescope} & \colhead{Filter} & \colhead{magnitude\tablenotemark{b}} \\
\colhead{(UT)} & \colhead{(days)} & \colhead{} & \colhead{} & \colhead{} 
}
\startdata
2005 April 16.4641 & 0.0025 &  Palomar 60-inch & $I$ & $18.82\pm 0.11$  \\
2005 April 16.4659 & 0.0043 &  Palomar 60-inch & $I$ & $18.86\pm 0.11$  \\
2005 April 16.4677 & 0.0061 &  Palomar 60-inch & $I$ & $19.16\pm 0.13$  \\
2005 April 16.4696 & 0.0080 &  Palomar 60-inch & $I$ & $19.35\pm 0.23$  \\
2005 April 16.4692 & 0.0076 & Palomar 200-inch & $K_s$ & $16.37\pm 0.21$ \\
2005 April 16.4714 & 0.0098 &  Palomar 60-inch & $I$ & $19.01\pm 0.12$  \\
2005 April 16.4731 & 0.0116 &  Palomar 60-inch & $I$ & $19.53\pm 0.18$  \\
2005 April 16.4772 & 0.0156 & Palomar 60-inch & $z'$ & $19.16\pm 0.27$\tablenotemark{c} \\
2005 April 16.4763 & 0.0147 & Palomar 200-inch & $K_s$ & $16.64\pm 0.20$ \\
2005 April 16.4833 & 0.0217 &  Palomar 60-inch & $I$ & $19.09\pm 0.20$  \\
2005 April 16.4828 & 0.0212 & Palomar 200-inch & $K_s$ & $16.79\pm 0.20$ \\
2005 April 16.4880 & 0.0264 &  Palomar 60-inch & $I$ & $19.34\pm 0.23$  \\
2005 April 16.4895 & 0.0279 & Palomar 200-inch & $K_s$ & $16.89\pm 0.24$ \\
2005 April 16.4960 & 0.0344 & Palomar 200-inch & $K_s$ & $16.99\pm 0.25$ \\
2005 April 16.5026 & 0.0410 & Palomar 200-inch & $K_s$ & $16.98\pm 0.29$ \\
2005 April 16.5089 & 0.0473 & ANU 2.3-meter & $R$ & $20.92\pm 0.05$ \\
2005 April 16.5502 & 0.0886 & ANU 2.3-meter & $R$ & $21.18\pm 0.12$ \\
2005 April 16.5543 & 0.0927 & ANU 2.3-meter & $R$ & $21.40\pm 0.15$ \\
2005 April 16.5585 & 0.0969 & ANU 2.3-meter & $R$ & $21.16\pm 0.08$ \\
2005 April 16.5626 & 0.1010 & ANU 2.3-meter & $R$ & $21.43\pm 0.11$ \\
2005 April 16.5669 & 0.1053 & ANU 2.3-meter & $R$ & $21.44\pm 0.26$ \\
2005 April 16.5702 & 0.1086 & ANU 2.3-meter & $R$ & $21.39\pm 0.16$ \\
2005 April 16.5744 & 0.1128 & ANU 2.3-meter & $R$ & $21.37\pm 0.16$ \\
2005 April 16.5779 & 0.1163 & ANU 2.3-meter & $R$ & $21.18\pm 0.17$ \\
2005 April 16.5816 & 0.1200 & ANU 2.3-meter & $R$ & $21.35\pm 0.16$ \\
2005 April 16.5850 & 0.1234 & ANU 2.3-meter & $R$ & $21.36\pm 0.10$ \\
2005 April 16.5893 & 0.1277 & ANU 2.3-meter & $R$ & $21.44\pm 0.27$ \\
2005 April 16.5924 & 0.1308 & ANU 2.3-meter & $R$ & $21.46\pm 0.22$ \\
2005 April 16.5975 & 0.1359 & ANU 2.3-meter & $R$ & $21.63\pm 0.34$ \\
2005 April 16.6092 & 0.1476 & ANU 2.3-meter & $R$ & $21.71\pm 0.30$ \\
2005 April 16.6187 & 0.1571 & ANU 2.3-meter & $R$ & $21.87\pm 0.46$ \\
2005 April 17.3895 & 0.9279 & Palomar 60-inch & $R$ & $22.62\pm 0.30$ \\
2005 April 17.4006 & 0.9390 & Palomar 60-inch & $z'$ & $< 20.38$\tablenotemark{c} \\
2005 April 17.4087 & 0.9467 &  Palomar 60-inch & $I$ & $22.05\pm 0.30$  \\
2005 April 18.2884 & 1.8264 & Palomar 60-inch & $R$ & $< 23.46$ \\ 
2005 April 18.4031 & 1.9411 &  Palomar 60-inch & $I$ & $< 22.83$        \\
2005 April 23.3511 & 6.8896 & Palomar 200-inch & $K_s$ & $< 19.70$ \\
\enddata
\tablenotetext{a}{Days since explosion have been calculated for the mid-point of each exposure.}
\tablenotetext{b}{Magnitudes have not been corrected for extinction. Limits are given as $3\sigma$.}
\tablenotetext{c}{AB system.}
\label{tab:NIRopt}
\end{deluxetable}

\clearpage

\begin{deluxetable}{lrcrccc}
\tablecaption{{\it HST/ACS} Observations of \grb}
\tablewidth{0pt} \tablehead{ 
\colhead{Date Obs} & \colhead{$\Delta t$} & 
\colhead{Exp. Time} & \colhead{Filter} & \colhead{{\it HST} mag\tablenotemark{a}} & \colhead{Johnson mag\tablenotemark{b}} \\
\colhead{(UT)} & \colhead{(days)} & \colhead{(sec)} & \colhead{} & \colhead{(AB)} & \colhead{(Vega)} \\ 
} 
\startdata 
2005 May 23.38 & 36.92 & 3282 & F775W & $24.35\pm 0.02$ & $23.82\pm 0.02$ \\
2005 May 23.46 & 37.00 & 3430 & F850LP & $23.83\pm 0.03$ & \nodata \\ 
2005 June 16.17 & 60.71 & 3986 & F775W & $25.88\pm 0.10$ & $25.36\pm 0.10$ \\
2005 July 11.15 & 85.69 & 4224 & F775W & $26.44\pm 0.22$ & $25.92\pm 0.22$ \\
2005 November 21.06 & 218.60 & 4224 & F850LP & \nodata & \nodata \\
2005 November 21.19 & 218.73 & 4224 & F775W & \nodata & \nodata \\
\enddata \tablenotetext{a}{AB system magnitudes in the {\it HST}
  filters given in column 4.  Photometry was done on residual images (see \S\ref{sec:hst}).
  We have assumed the source flux to be negligible in the final
  (template) epoch.  Magnitudes have not been corrected for extinction.}
\tablenotetext{b}{We convert the F775W magnitudes in column 5 to Johnson $I$-band (Vega system) as described in \S\ref{sec:hst}.
  Magnitudes have not been corrected for extinction.}
\label{tab:hst}
\end{deluxetable}

\clearpage

\begin{deluxetable}{lrrrc}
\tablecaption{Spectroscopic lines for XRF\,050416a}
\tablewidth{0pt} \tablehead{
\colhead{Line} & \colhead{$\lambda$ (rest)} & \colhead{$\lambda$ (observed)\tablenotemark{a}} & \colhead{Redshift} & \colhead{Flux\tablenotemark{b}} \\
\colhead{} & \colhead{(\AA)} & \colhead{(\AA)} & \colhead{} & \colhead{($\times 10^{-17}\rm erg~cm^{-2}~s^{-1}$)}
}
\startdata
$\rm [OII]$ & 3728.38   &    6162.08    &   0.6528 &  $9.6\pm 0.4$ \\
H$\gamma$ & 4341.72 & 7174.64 &  0.6525 & $1.2\pm 0.2$ \\
H$\beta$  & 4862.72   &   8037.14    &     0.6528    &      $3.7\pm 0.5$  \\
$\rm [OIII]$  & 4960.30 &      8198.79    &     0.6529   &    $2.5\pm 0.3$  \\
$\rm [OIII]$  & 5008.24   &      8277.57    &  0.6528      &      $8.4\pm 0.4$  \\
\enddata
\tablenotetext{a}{Observed wavelengths have been corrected to vacuum.}
\tablenotetext{b}{Flux values have not been corrected for Galactic extinction}.
\label{tab:lines}
\end{deluxetable}

\clearpage

\begin{deluxetable}{lrrrr}
\tablecaption{Radio Observations of XRF\,050416a}
\tablewidth{0pt} \tablehead{
\colhead{Date Obs} & \colhead{$\Delta t$} & \colhead{$F_{\nu,1.43~\rm GHz}$\tablenotemark{a}} & \colhead{$F_{\nu,4.86~\rm GHz}$} & \colhead{$F_{\nu,8.46~\rm GHz}$} \\
\colhead{(UT)} & \colhead{(days)} & \colhead{($\mu$Jy)} & \colhead{($\mu$Jy)} & \colhead{($\mu$Jy)}
}
\startdata
2005 April 16.49 & 0.026 & \nodata & \nodata & $20\pm 51$ \\
2005 April 22.03 & 5.57  & \nodata & $200\pm 46$ & $101\pm 34$ \\
2005 April 28.29 & 11.83 & \nodata & $188\pm 42$ & $132\pm 30$ \\
2005 May 1.35    & 14.89 & \nodata & $201\pm 43$ & \nodata \\
2005 May 31.10   & 44.64 & \nodata & \nodata & $431\pm 46$ \\
2005 June 3.99   & 48.53 & \nodata & $585\pm 48$ & \nodata \\
2005 June 8.03   & 52.57 & $420\pm 190$ & $562\pm 51$ & $398\pm 33$ \\
2005 June 16.97  & 61.51 & \nodata & $505\pm 49$ & $330\pm 35$ \\ 
2005 June 20.16  & 64.70 & \nodata & \nodata & $286\pm 33$ \\
2005 July 1.92   & 76.46 & $843\pm 214$ & $321\pm 48$ & $189\pm 36$ \\
2005 July 30.02  & 104.56 & $0\pm 192$ & $68\pm 36$ & $93\pm 41$ \\
2005 August 14.97 & 120.51 & \nodata & \nodata & $72\pm 42$ \\
2005 August 21.89 & 127.42 & \nodata & \nodata & $41\pm 47$ \\
2005 September 19.90 & 156.44 & \nodata & $0\pm 54$  & \nodata\\
2005 October 15.73 & 182.27 & \nodata & $33\pm 47$ & $89\pm 38$ \\
\enddata
\tablenotetext{a}{All errors are given as $1\sigma$ (rms).}  
\label{tab:vla}
\end{deluxetable}

\clearpage

\clearpage

\clearpage

\clearpage

\clearpage

\clearpage

\clearpage

\end{document}